%% file: manuscript.tex
\documentclass[lettersize,journal]{IEEEtran}
\usepackage{amsmath,amsfonts}
\usepackage{algorithmic}
\usepackage{array}
\usepackage{cite}
\usepackage[caption=false,font=normalsize,labelfont=sf,textfont=sf]{subfig}
\usepackage{textcomp}
\usepackage{stfloats}
\usepackage{url}
\usepackage{verbatim}
\usepackage{graphicx}
\usepackage{adjustbox}
\def\BibTeX{{\rm B\kern-.05em{\sc i\kern-.025em b}\kern-.08em
    T\kern-.1667em\lower.7ex\hbox{E}\kern-.125emX}}
\usepackage{balance}

\usepackage{url,hyperref,lineno,microtype,subcaption}
\usepackage{siunitx}
\usepackage{upgreek}
\usepackage{seqsplit}
\usepackage{booktabs}

\DeclareSIUnit\Molar{\textsc{M}}
\DeclareSIUnit\Siemens{\textsc{S}}

\begin{document}
\title{Neural and Cognitive Impacts of AI: The Influence of Task Subjectivity on Human-LLM Collaboration}
\author{Matthew Russell$^{1}$, Aman Shah$^{1}$, Giles Blaney$^{1}$, Judith Amores$^{2}$, Mary Czerwinski$^{3}$, and Robert J.K. Jacob$^{1}$\\$^{1}$Tufts University, Medford, Massachusetts, USA\\
$^{2}$Microsoft Research, Cambridge, Massachusetts, USA\\
$^{3}$Microsoft Research, Redmond, Washington, USA}

\maketitle

\noindent © 2025 IEEE. Personal use of this material is permitted. Permission
from IEEE must be obtained for all other uses, in any current or future
media, including reprinting/republishing this material for advertising or
promotional purposes, creating new collective works, for resale or
redistribution to servers or lists, or reuse of any copyrighted
component of this work in other works.\\

\begin{abstract}
AI-based interactive assistants are advancing human-augmenting technology, yet their effects on users' mental and physiological states remain under-explored. We address this gap by analyzing how Copilot for Microsoft Word, a LLM-based assistant, impacts users. Using tasks ranging from objective (SAT reading comprehension) to subjective (personal reflection), and with measurements including fNIRS, Empatica E4, NASA-TLX, and questionnaires, we measure Copilot's effects on users. We also evaluate users' performance with and without Copilot across tasks. In objective tasks, participants reported a reduction of workload and an increase in enjoyment, which was paired with objective performance increases. Participants reported reduced workload and increased enjoyment with no change in performance in a creative poetry writing task. However, no benefits due to Copilot use were reported in a highly subjective self-reflection task. Although no physiological changes were recorded due to Copilot use, task-dependent differences in prefrontal cortex activation offer complementary insights into the cognitive processes associated with successful and unsuccessful human-AI collaboration. These findings suggest that AI assistants' effectiveness varies with task type—particularly showing decreased usefulness in tasks that engage episodic memory—and presents a brain-network based hypothesis of human-AI collaboration. 
\end{abstract}

\begin{IEEEkeywords}
Large Language Model, Human-Computer Interaction, Brain-Computer Interfaces, functional Near-Infrared Spectroscopy, Empatica, Copilot
\end{IEEEkeywords}

\section{Introduction}
\IEEEPARstart{B}{y} giving humans new ways to access information, Large Language Model (LLM) based interactive assistants such as ChatGPT promise to revolutionize the way we work. Indeed, considering the significant mental demands of complex creative and decision-making tasks, the LLM-based assistant could represent a paradigm shift in the cognitive landscape of human users. However, little is known about the effects such systems actually have on their users. Does the user disengage and let the assistant do all the work? Do they engage more? Do they produce better or worse outputs? More generally, what effects do such tools have on users? How can this inform the design of future interactive LLM-based assistants? Are there specific aspects of human neural function which correspond to beneficial or poor experience while working with LLM tools? In this work, we explore these questions with a variety of measurement techniques to investigate the effects of using an LLM on a user's self-reported and physiological mental workload and stress, as well as their objective performance as they perform an array of different tasks intended to target different aspects of human experience. 
\\\\
For the LLM-based assistant in our study we used a version of Microsoft Word in the Microsoft 365 suite equipped with the Microsoft Copilot interactive Artificial Intelligence (AI) assistant. We used a 4 x 2 within-subjects design in which each of our four tasks had two equally difficult variants; for each task, participants did one with and one without the Copilot assistant. Experimental tasks were defined along a gradient of subjectivity estimated to interface with different aspects of human experience and correspond to the degree of `difficulty' for the assistant. To quantitatively measure mental workload we employed both physiological and self-report methods. Physiological measures include the use of functional Near-Infrared Spectroscopy (fNIRS) to measure changes prefrontal cortex hemoglobin concentration and the Empatica E4 device to observe Heart Rate (HR), Heart Rate Variability (HRV) and Electrodermal Activity (EDA). Physiological measures are complemented by quantitative self-reported data from both the NASA Task Load Index (NASA-TLX) and questionnaires, and qualitative analysis via user feedback is also performed. Quality of the users' performance with and without the \texttt{AI} assistant was also assessed.

\section{Background and Related Work}
\subsection{Large-Language Models and HCI}
Human-computer interaction research on user impact from LLMs is still in its beginning stages. Much of the current research is still based in analyzing user output and using qualitative methods to understand user preferences \cite{ChangWang24, Yuan22}. However, in recent years, quantitative methods have played a more significant role with studies looking at how user performance and time spent on a task changes with the use of LLMs \cite{Dellacqua23, NoyZhang23}. Notable areas of application where research has been conducted to understand the effects LLM tools have on users include writing, computer programming, and decision making. 

\subsubsection{Writing}
Yuan tested an LLM story writing tool with professional authors to gain insights into the effectiveness of LLMs in supporting creative writing \cite{Yuan22}. Nihil \cite{Nikhil2022writing} examined the potential and challenges of LLM use for creative writing, and Reza produced ABScribe, a novel interface for more easily integrating human and machine-generated work in Human-AI co-writing tasks \cite{RezaABScribe}. Other researchers have explored whether there is a difference between quality in AI and human-generated literary short texts \cite{Gunser22}. Both Yuan and other studies have, using both qualitative and quantitative methods, demonstrated a productivity boost when using LLMs for work-related tasks, especially for novice and low-skilled workers \cite{Brynjolfsson23, Dellacqua23, Yuan22}. However, the complexity of these systems reduces their benefits for novice users who don't know how to use them effectively, especially in light of the sophistication required for prompt design \cite{PratherReeves23, Zamfirescu23}.  

\subsubsection{Programming}
Computer programming has also proven an effective testing ground for studying the effects of LLM tools on users. Ziegler \cite{Ziegler2023productivity} performed a comprehensive study investigating the effects of Github Copilot on users, with a specific interest in productivity while Nguyen studied the challenges that non-expert users face when using LLM-based tools to assist in programming \cite{nguyen2023LLM}.

\subsubsection{Decision Making}
Researchers have also investigated the benefits, drawbacks, and limitations of using LLM tools as an integral component of decision making processes. Lawless investigated the combination of LLMs with Constraint Programming to facilitate decision making \cite{LawlessDecisionSupport}. Chiang studied the use of AI tools to help decision making specifically in group-based settings \cite{ChiangDecision}. Bu\c{c}inca has studied intrinsic motivation in Human-AI decision making, and Lakkaraju investigated the fairness and efficacy of LLM tools used in the context of financial decision making \cite{LakkarajuFinancial}. 

\subsubsection{Other LLM-based User Studies}
Other avenues of approach for investigating the effects of LLM-based tools on users include Arakawa's work on adapting an LLM chatbot towards executive coaching \cite{arakawaChatbot}, Huang's work exploring the use of LLM assistants to help prevent driver fatigue \cite{HuangDriver}, Suh's work on LLM-based tools for structured design space exploration \cite{SuhDesignSpace} and multilevel sensemaking \cite{SuhSensemaking}, and Tankelevitch's work on mapping the underlying metacognitive load while using AI tools \cite{TankelevitchMetacognitive}.

\subsection{fNIRS and the Prefrontal Cortex}

\subsubsection{fNIRS} fNIRS uses diffuse optical imaging of near-infrared light to non-invasively measure changes in oxygenated $\Delta$[HbO] and deoxygenated $\Delta$[Hb] hemoglobin concentrations in the human brain \cite{fantini2020frequency}. These measures can be connected to changes in cerebral blood flow, which, in turn, are connected to brain oxygen demand and, thus, functional activation.

\subsubsection{The Prefrontal Cortex (PFC)} Activation in the PFC, especially the anterior and dorsolateral structures, is associated with a wide variety of cognitive tasks including problem-solving, planning, reasoning, and working memory \cite{Bunce11, Koechlin99, Ramnani04}. Research in this area has utilized a variety of functional neuroimaging tools, including functional magnetic resonance imaging \cite{Desposito00, Manoach97} and fNIRS \cite{vermeij2014very}. This multimodal research has also elucidated that the many cognitive functions located in or supported by the PFC provide the cognitive flexibility necessary for creative processing and thinking \cite{Dietrich04, Shah13}. This substantial association allows for the use of prefrontal cortex activation as a measurement of user mental workload when completing a variety of tasks.

\subsubsection{HCI with PFC and fNIRS} In particular, studies with air traffic control operators and others have shown that fNIRS is particularly useful in assessing mental workload as users complete ecologically valid tasks on an interface \cite{Ayaz12}. Even more, research has shown the utility of fNIRS in classifying high and low levels of mental workload, allowing for evaluation of interfaces based on their impact on a user's cognitive load \cite{Bosworth19, Girouard09}. Indeed, Hirshfield et al. \cite{Hirshfield11} shows that fNIRS enhances usability testing because it provides quantitative information on the cognitive demands an interfaces places on a user. 

\subsubsection{Very Low Frequency Oscillations (VLF) with fNIRS and the PFC}
Research in fMRI and fNIRS has highlighted the accessibility and usefulness of observing Low Frequency (LF) [0.07 Hz - 0.2 Hz] and Very Low Frequency (VLF) [0.02 Hz - 0.07 Hz] oscillations as correlates of cerebral hemodynamics \cite{obrig2000spontaneous, Sassaroli2012}. In particular, a decrease in the VLF band has been shown to relate to task-based cortical activation \cite{obrig2000spontaneous}. Further, such task-based cortical activation in the prefrontal cortex has been detected with fNIRS \cite{vermeij2014very}.

\begin{figure}
    \centering
    \includegraphics[width=.9\linewidth]{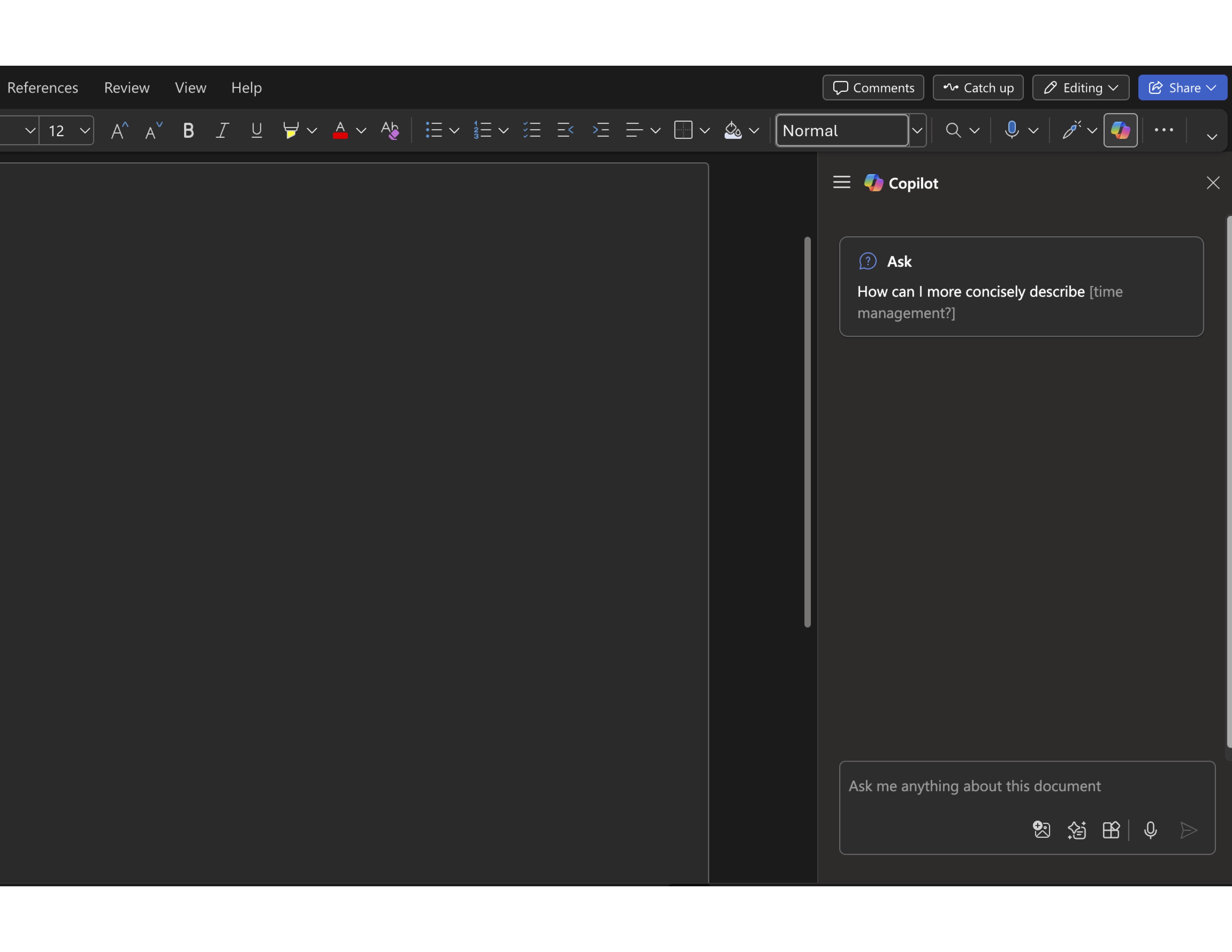}
    \caption{Microsoft Word with the integrated Copilot sidebar}
    \label{fig:Copilot}
\end{figure}

\subsection{Microsoft Copilot}
Copilot is an extension of the standard Microsoft Word interface which leverages AI to assist users throughout a variety of tasks. Although the Copilot ecosystem in Word allows users a wide array of functionality through multiple contexts, in order to minimize training time for our users as well as the potential for interface-based confounds, we focused the user's interaction with Copilot to a single chat window on the side of the Word screen (see Figure \ref{fig:Copilot}). This chat allows users to interact with the Copilot assistant, and it in turn interfaces with a LLM to produce relevant responses. While the specifics of which LLM is used are abstracted from the Word interface, Microsoft's documentation specifies that it leverages a variant of GPT-4 along with the text-to-image model DALL-E \cite{MicrosoftCopilotOverview}. For this research, the relevant tasks that Copilot can perform are: text generation and refinement, answering queries related to the current document, or queries requesting general information or answers to specific questions. 


\subsection{Empatica E4}
The Empatica E4 device is a wristwatch-like device that measures Photoplethysmogram (PPG) and Electrodermal activity (EDA). From PPG, it produces measurements of Heart Rate (HR) and Inter-Beat-Interval (IBI) data, which can be used to determine Heart Rate Variability (HRV) \cite{milstein2020validating}. Among the Empatica E4's measurements of \texttt{HR}, \texttt{IBI}, and \texttt{EDA}, \texttt{HR} has been shown to be the most reliable in comparison to gold-standard methods \cite{Schuurmans2020}. And, although \texttt{EDA} and \texttt{IBI} have not performed as well against baseline benchmarks, particularly in collection settings separate from rest \cite{milstein2020validating}, the E4 has been widely used by researchers across disciplines to measure affect \cite{ShenBa2023, Schmidt2019} and stress \cite{Hickey2021, kim2020development}. 

\section{Research Questions}
The primary aim of this study is to explore the effects of using an interactive LLM system (in this case, Copilot for Microsoft Word) on human users. Our specific research questions follow below. For each question \texttt{RQX} we are interested in \texttt{RQX-A}: overall effect, \texttt{RQX-B}: effects within each task, and \texttt{RQX-C}: effects that differ along the gradient of subjectivity.

\subsubsection*{RQ1} \label{RQ1} Does the use of the Copilot assistant change users' workload levels as measured by \texttt{NASA-TLX}? 
\subsubsection*{RQ2} \label{RQ2} Does using the Copilot assistant change users' levels of prefrontal cortex activation as measured by fNIRS? 
\subsubsection*{RQ3} \label{RQ3} Does the use of the Copilot assistant change users' levels of stress as measured by Heart Rate (\texttt{HR}), Heart Rate Variability (\texttt{HRV}), and Electro Dermal Activity (\texttt{EDA})?
\subsubsection*{RQ4} \label{RQ4} Does using the Copilot assistant change the quality of users' output?
\subsubsection*{RQ5} \label{RQ5} How do users feel about using the Copilot assistant?

\section{Materials and Methods}

\subsection{Study Tasks}
We modeled our tasks along a \textit{gradient of subjectivity}. We designed this gradient along theoretical considerations of neurological systems, and developed tasks with practical experimental constraints in mind. At one end of the gradient are highly structured tasks with objectively clear and correct answers: we hypothesized that these tasks would engage participants in mental workload typically associated with prefrontal cortex activity; we expected these tasks would allow Copilot to meaningfully assist users, and would result in a corresponding decrease of prefrontal activation relating to decreased workload. At the opposite end of the gradient are open-ended tasks with highly subjective elements: we hypothesized that these tasks would engage participants in prefrontal activation associated with episodic memory; we expected that these tasks would present significant challenges for the AI assistant and would not affect brain function.

Determining the specific tasks that we would have our users engage in required much care and several iterations to strike a balance between tasks that were easy enough for the LLM that it could perform them perfectly with a single click and tasks that were too lengthy and involved for users to accomplish in a reasonable amount of time. A particular challenge we discovered from prior research and our own tests is that large language models are most effective in tasks with high complexity and low ambiguity \cite{Haslberger23}; that is, Copilot produces highly detailed and effective output in direct proportion to the level of detail and structure of the task: the more structured and detailed the task, the more structured and detailed the output from Copilot. After iterative refinement, we settled on a set of four task groups: \textbf{reading comprehension} (objective, fact-based, requires working memory), \textbf{event planning} (structured, but creative), \textbf{poetry writing} (creative with personal elements), and \textbf{personal reflection} (highly subjective and directly connected to personal experience and episodic memory). For each task type, we created two subtasks designed to be equally difficult. The full text of the tasks themselves, and a statistical analysis testing mental workload changes between the subtasks (no significant differences were found), can be found in the supplementary material. 

\subsubsection{\textbf{Reading Comprehension}} These questions were slightly modified versions of examples taken from the CollegeBoard's Scholastic Aptitude Test (SAT), and were easily answered by Copilot. This task served as a baseline, representing highly objective problem-solving with minimal subjectivity. We anticipated standard cognitive demands on users without Copilot, and minimal cognitive demand when assisted by the LLM. 

\subsubsection{\textbf{Event Planning}} These tasks asked the user to design and plan an event with structured and detailed to-do checklists of the event-related information. While still structured, these tasks were more open-ended than those in \texttt{SAT}, and required more subjective, personal, and creative input. We hypothesized that Copilot would be helpful to the user in completing this task, but that it would require more work from users in the Copilot condition as compared to \texttt{SAT}.

\subsubsection{\textbf{Poetry Writing}} These tasks asked the user to write a short poem of 10-15 lines on a broad theme such as joy or nature. This task represents a substantial shift toward subjective material, requiring purely creative expression that, at least in the without-LLM assistance condition, would necessarily draw on subjective personal experience. We believed that this task would engage more fully with the episodic memory than the first two, and that the LLM assistant would enable users to quickly produce output, but that it also would struggle to assist them given the inherently subjective nature of a poem.

\subsubsection{\textbf{Personal Reflection}} These tasks asked the user to reflect on their favorite album or movie and discuss why it was their favorite based on their personal experiences. This task was designed to maximally engage purely subjective, autobiographical episodic memory; we therefore hypothesized that it would be quite challenging for the LLM tool to meaningfully assist the user during these tasks. 

\subsection{Study Structure}
All users signed informed consent documents prior to beginning the study, which was approved by the Tufts University Institutional Review Board. We provided an initial survey regarding familiarity with AI tools. Participants then did a 5 minute training task to familiarize them with the Copilot assistant. This included a variety of prompts for the user to use with the assistant to better help them understand what it could and could not do. Participants were able to ask questions prior to beginning the tasks if they needed help with Copilot. Users then completed each of the four tasks in a randomized order counterbalanced across participants. The choice of which subtask would be completed with the LLM assistant was likewise counterbalanced. After each task participants filled out post-task surveys including the NASA-TLX, a space for users to write any comments they would like, and a follow-up question rated on a scale of [0-10]: ``How would you rate your overall experience with this task? (0=Terrible, 10=Amazing)''. The participants were compensated with an Amazon gift card (\$25) for their time. 

\subsection{Data Collection and Preprocessing}
\subsubsection{Demographics} We recruited 20 healthy individuals (7 Male, 10 Female, 3 opted not to disclose) for the study, ranging from 18-25 years old (mean 21). 


\subsubsection{Exclusions from Physiological Data}
Four participants were excluded due to excessive noise across multiple trials seen through visual inspection of the fNIRS data. One fNIRS participant was excluded because an experimenter incorrectly marked the data and one was excluded because the user refused to wear the fNIRS headband. Within otherwise used fNIRS data, frequency domain $\Delta$[HbD]$\upphi$ data of the left prefrontal cortex for two participants in one task session exceeded 1.5 times the Interquartile Range (IQR) across all participants: the data were also excluded. From the Empatica data, three users had invalid signal connection issues between the E4 and our collection device during collection time (Google Pixel 6 Phone), two users were excluded due to manual marker input errors, and one user declined to wear the wristband.  

\subsubsection{fNIRS}
\begin{figure}
    \centering
    \includegraphics[width=.6\linewidth]{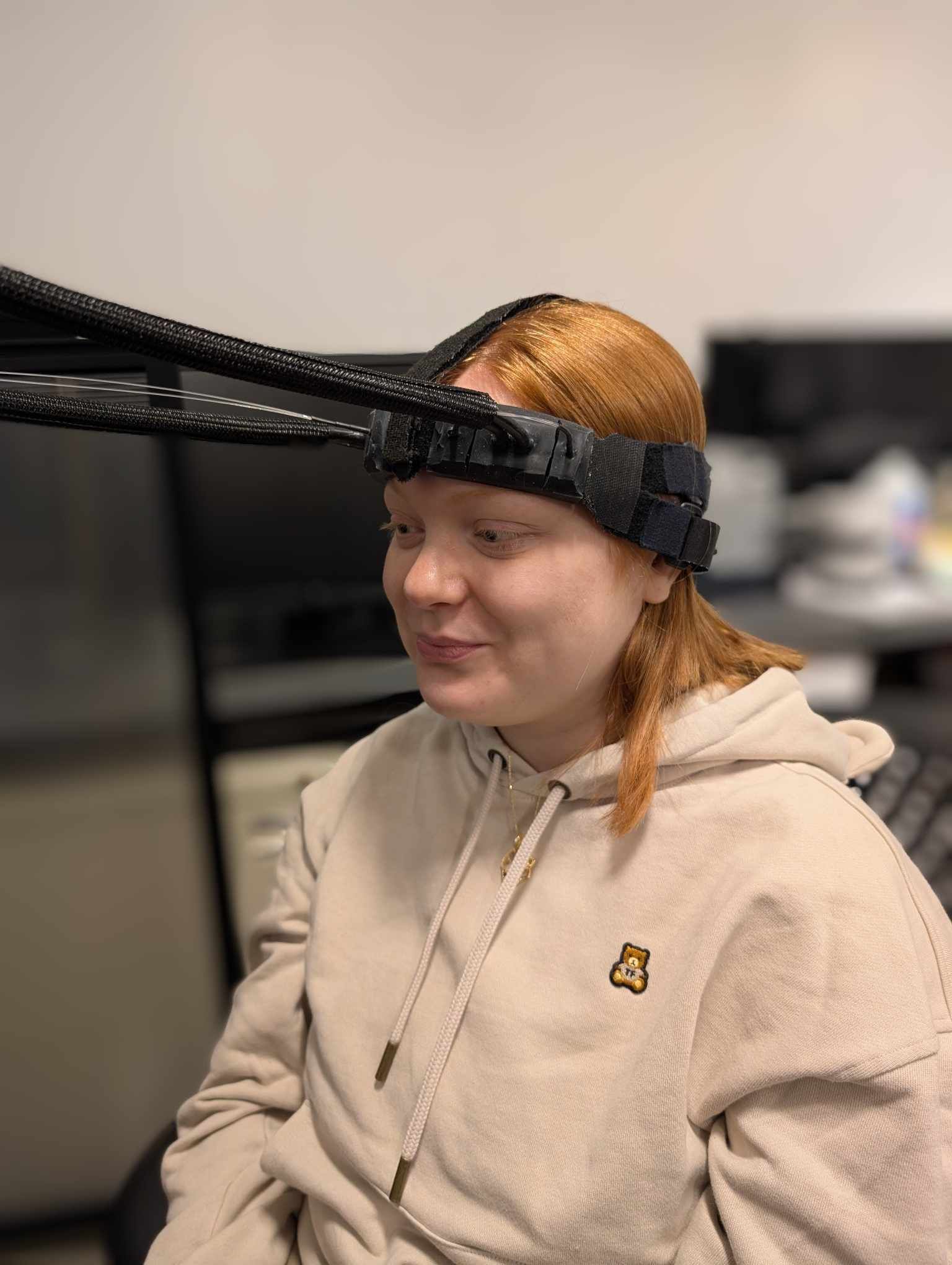}
    \caption{User wearing a functional near-infrared spectroscopy (fNIRS) device.}
    \label{fig:fNIRS}
\end{figure}

We utilized a frequency-domain near-infrared spectroscopy device (ISS Imagent, Champaign, IL USA) operating at a modulation frequency of 110~MHz and with wavelengths of 690~\si{\nano\meter} and 830~\si{\nano\meter}. Two custom-made optical probes were placed on the subject’s forehead, one each for the left and right hemispheres. The probes were secured to the subject’s head using an adjustable loop headband which passed through the center of the probes. Centroids of the probes were located over the prefrontal cortex of the associated probe (Figure \ref{fig:fNIRS}) at the approximate locations of AF7 and AF8 in the Standard 10-10 Electrode Configuration \cite{koessler2009automated}. Each optical probe had optode geometry designed for the dual-slope (DS) method \cite{blaney2020phase}. Each probe consisted of two source positions, each with two wavelengths and two detectors. For each DS probe, data from all combinations of sources and detectors were collected, resulting in a total of four single-distance (SD) measurements (source-detector distances ($\rho$): two of 25 mm and two of 35 mm) each of frequency-domain intensity amplitude (I) and phase ($\upphi$) \cite{WangHuang21}. The light was delivered to each probe via 400~\si{\micro\meter} diameter multi-mode fibers and collected by 5~\si{\milli\meter} diameter fiber bundles. These fibers were held in-place by a flexible plastic mesh and were encapsulated in black silicone. 
\\\\
Data collection occurred in BOXY, a software provided ISS Imagent. Nominal gains for each detector were found using BOXY for each user prior to beginning the study. The I and $\upphi$ data for each source-detector pair was processed using DS methods, resulting in measurements of $\Delta$[HbO] ($\upmu$M) and $\Delta$[HbR] ($\upmu$M) for each of I and $\upphi$ \cite{blaney2020phase}. Baseline correction for each trial was performed with the initial 15 seconds for each trial, and the last 15 seconds of each trial were discarded. Each measurement for each trial was linearly detrended, and a 5th order Butterworth bandpass filter was applied of the range [0.02, 0.2]~Hz \cite{klein2019signal}. For statistical analysis, $\Delta$[HbD] was calculated by $\Delta$[HbO]-$\Delta$[HbR] \cite{Kreplin2013}, frequency domain transformation was performed using the Multitaper method \cite{percival1993spectral, candy2019multitaper}, Simpson's rule was used to integrate over the VLF frequency band \cite{fdez2021cross}, and the resulting values were log-transformed. Statistical analyses were then performed on the \texttt{DSI} and \texttt{DS$\upphi$} data \cite{obrig2000spontaneous, vermeij2014very}, with separate models created for each probe and measurement value. For convenience, we refer to the log total power in the VLFO band of the fNIRS signal as \texttt{fNIRS} in the text below. Note that although we do not use short channels for artifact removal, the DS method leverages counter-posing pairs of channels to perform removal of extracerebral information, including movement artifacts and scalp hemodynamics \cite{blaney2020phase}.
\\
\subsubsection{Empatica E4} Our preprocessing steps for the various empatica streams was as follows.

\begin{itemize}[]
\item [\texttt{HR}] We extracted the mean \texttt{HR} for each trial. 
\item [\texttt{HRV}] Because Empatica's inter-beat-interval recording has preprocessing of the signal applied in advance of the point of measurement from the device that removes most of the non-normal beats in the RR interval, we used Empatica's \texttt{IBI} to represent the \texttt{IBI} of normal sinus beats (\texttt{NN} \cite{Shaffer2017}) \cite{Schuurmans2020}, and used the standard deviation of the Empatica \texttt{IBI} data as \texttt{SDNN} for our \texttt{HRV} calculation. We excluded trials with an \texttt{IBI} value outside of the range [1, 125]~ms (5/81 trials were excluded). 
\item [\texttt{EDA}] \label{EDASymp} Each of the trials were bandpass filtered with a 4th order Butterworth filter of the range [0.01, 0.8]~Hz \cite{PosadaQuintero2016}. We then transformed the signal into the frequency domain using the same process as with the fNIRS data; the frequency band extracted was [0.045 0.25]~Hz which has been shown to produce a reliable inference of sympathetic \texttt{EDA} \cite{PosadaQuintero2016}.\\
\end{itemize}

\subsubsection{NASA-TLX}
We produced unweighted average TLX score for each participant's response for each condition \cite{grier2015high}.
\\
\subsubsection{Task Evaluation Scores}
\label{subsubsection:task_evaluation}
For SAT, quality scores were simply defined as the percent of correct answers total per task. For the other three tasks we had three members of our research team grade each of the submissions provided for the \texttt{PLANNING}, \texttt{POEM} and \texttt{REFLECTION} tasks independently, rating each submission on a [1-5] scale for both of \textit{breadth} and \textit{depth}. Consistency of the graders' output was measured with Intraclass Correlation \texttt{ICC} \cite{bartko1966intraclass}, specifically using a two-way mixed-effects model considering consistency over the mean of k raters (\texttt{ICC3k}) \cite{Koo2016}. Quality scores for \textit{breadth} and \textit{depth} were averaged across graders, and the resulting scores were then averaged to produce a single score value for each user for each task. Quality scores for each task were then normalized across users to a 0-1 scale. 

\subsection{Statistical Methods}
To account for the repeated measures design of our study we analyzed our data using Linear Mixed Models (LMMs) \cite{oberg2007linear}. We created separate models for each research question using the following \texttt{R} formula as a template:
\begin{equation}
\label{formula:initial}
    \texttt{DV} \sim \texttt{CONDITION} * \texttt{TASK} + \texttt{(1}|\texttt{PID/TASK)}
\end{equation}
Where \texttt{DV} represents the measured dependent variable of interest (\texttt{TLX}, \texttt{fNIRS}, \texttt{HR}, \texttt{HRV}, \texttt{IBI}, \texttt{PERFORMANCE}, or \texttt{ENJOYMENT}), \texttt{CONDITION} is a factor with two levels indicating use of Copilot (with-Copilot (\texttt{AI}) or without-Copilot (\texttt{NAI})), and \texttt{TASK} is a factor with four levels indicating the type of task performed (\texttt{SAT}, \texttt{PLANNING}, \texttt{POEM}, or \texttt{REFLECTION}). Random intercepts are specified for each participant (\texttt{PID}), with nested intercepts within participant for each \texttt{TASK}. For each model, likelihood ratio tests (LRTs) were used to refine the random effects structure \cite{Vuong1989}; models that showed better fit without the nested random effect of \texttt{TASK} within \texttt{PID} had this term removed. 

ANOVA results from the LMMs for \texttt{CONDITION} are used to determine significance for all RQX-A. If interaction of \texttt{CONDITION} $\times$ \texttt{TASK} demonstrates significance, post-hoc contrasts are performed among the emmeans for \texttt{CONDITION} within levels of \texttt{TASK} to answer all RQX-B. To answer all RQX-C questions respective of Copilot (done if \texttt{CONDITION} $\times$ \texttt{TASK} is significant), custom emmeans contrasts are performed to test the effect of \texttt{CONDITION} across different pairs of \texttt{TASK} levels. 

To answer all RQX-C questions irrespective of Copilot (done if \texttt{CONDITION} $\times$ \texttt{TASK} is not significant, but \texttt{TASK} is), post-hoc contrasts are performed among the emmeans comparing levels of \texttt{TASK}. 

For all tests, $\alpha$ is set at 0.05, except in the case of omnibus testing for fNIRS data and empatica data, where we apply Bonferroni correction; for fNIRS, we consider the measures of \texttt{DSI} and \texttt{DS$\upphi$} for each side of \texttt{L} and \texttt{R} as related, and thus $\alpha$ is adjusted to 0.025; for Empatica, we consider \texttt{HR} and \texttt{HRV} related, so $\alpha$ is adjusted to 0.025 for those tests.

For effect sizes, we report partial Epsilon squared ($\epsilon_p^2$), also known as adjusted partial eta squared (adj. $\eta_p^2$), which quantifies the proportion of variance associated with a given effect while controlling for other variables in the model, and reduces the bias introduced by the usual $\eta_p^2$ calculation \cite{Mordkoff2019}\footnote{Despite that it is less often reported than $\omega_p^2$ it has been shown that $\epsilon_p^2$ is less biased \cite{Carroll1975}.}. 

\subsection{Software Tools}
Data were processed in the \texttt{Python} programming language. The \texttt{pandas} \cite{reback2020pandas} and \texttt{numpy} \cite{harris2020array} libraries were used for data aggregation and filtering. Multitaper frequency transformations were performed with the \texttt{mne} package \cite{GramfortEtAl2013a}. The \texttt{rpy2} package was used to run \texttt{R} code, wherein we did all statistical analyses. \texttt{lmerTest} was used to create the Mixed-Effects Regression models \cite{lmerTest}, estimated marginal means and associated pairwise comparisons were calculated with the \texttt{emmeans} package \cite{emmeans}. Effect size calculations and associated confidence intervals were determined with the \texttt{effectsize} package \cite{Ben-Shachar2020}. Visualizations and associated error-bar calculations were made with the \texttt{Seaborn} \cite{Waskom2021} package, and error-bars represent 95\% confidence levels using a 10,000 sample multilevel bootstrap grouped by participant id to accounting for repeated measures within participants \cite{efron1979bootstrap, Waskom2021}.

\section{Results}

\subsection{\nameref{RQ1}: TLX Workload Results} 

\subsubsection{\nameref{RQ1}-A Results}
\begin{figure}[h!]
    \centering
    \includegraphics[width=\linewidth]{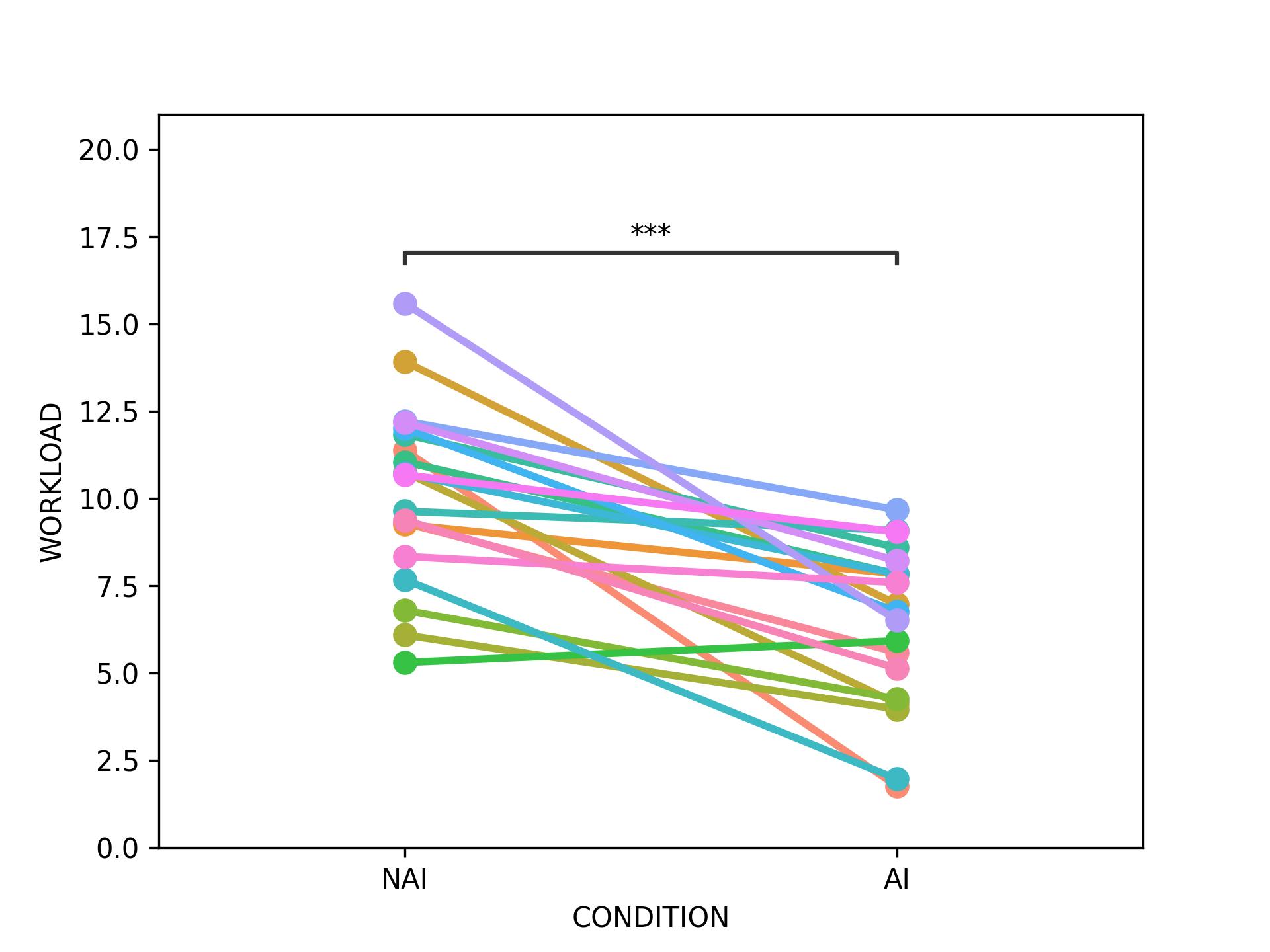}
    \caption{\texttt{TLX} scores in the \texttt{NAI} (without Copilot) and \texttt{AI} (with Copilot) conditions over all tasks. Each line represents a unique user. Self-reported workload generally decreased when using Copilot. Further discussion of separate effects across levels of \texttt{TASK} is below. }
    \label{fig:CONDITION}
\end{figure}

The Copilot condition resulted in overall lower \texttt{TLX} scores ($F_{1, 133}=60.42, p<0.001, \epsilon_{p}^{2}=.31$). Results are visible in Table \ref{tab:ANOVA} and visualized in Figure \ref{fig:CONDITION}. 

\begin{table}[h!]
\centering
\caption{ANOVA result from a model with \texttt{WORKLOAD} as the \texttt{DV} in Formula \ref{formula:initial}. Although overall self-reported workload decreased with Copilot, differences were found with an interaction with \texttt{CONDITION}.}
\label{tab:ANOVA}
\begin{adjustbox}{width=\linewidth}

\begin{tabular}{lrrrrccc}
\toprule
\textbf{Factor} & \textbf{df1} & \textbf{df2} & \textbf{F} & \textbf{p} & \textbf{sig.} & \textbf{$\epsilon_{p}^{2}$} & \textbf{$\epsilon_{p}^{2}$ CI}\\
\midrule
CONDITION & 1 & 133 & 60.42 & $<$0.001 & *** & 0.31 & [0.20,0.40]\\
TASK & 3 & 133 & 9 & $<$0.001 & *** & 0.15 & [0.06,0.23] \\
CONDITION $\times$ TASK & 3 & 133 & 7.07 & $<$0.001 & *** & 0.12 & [0.03,0.20] \\
\bottomrule
\end{tabular}
\end{adjustbox}
\end{table}

\subsubsection{\nameref{RQ1}-B Results}
\begin{figure}[h!]
    \centering
    \includegraphics[width=\linewidth]{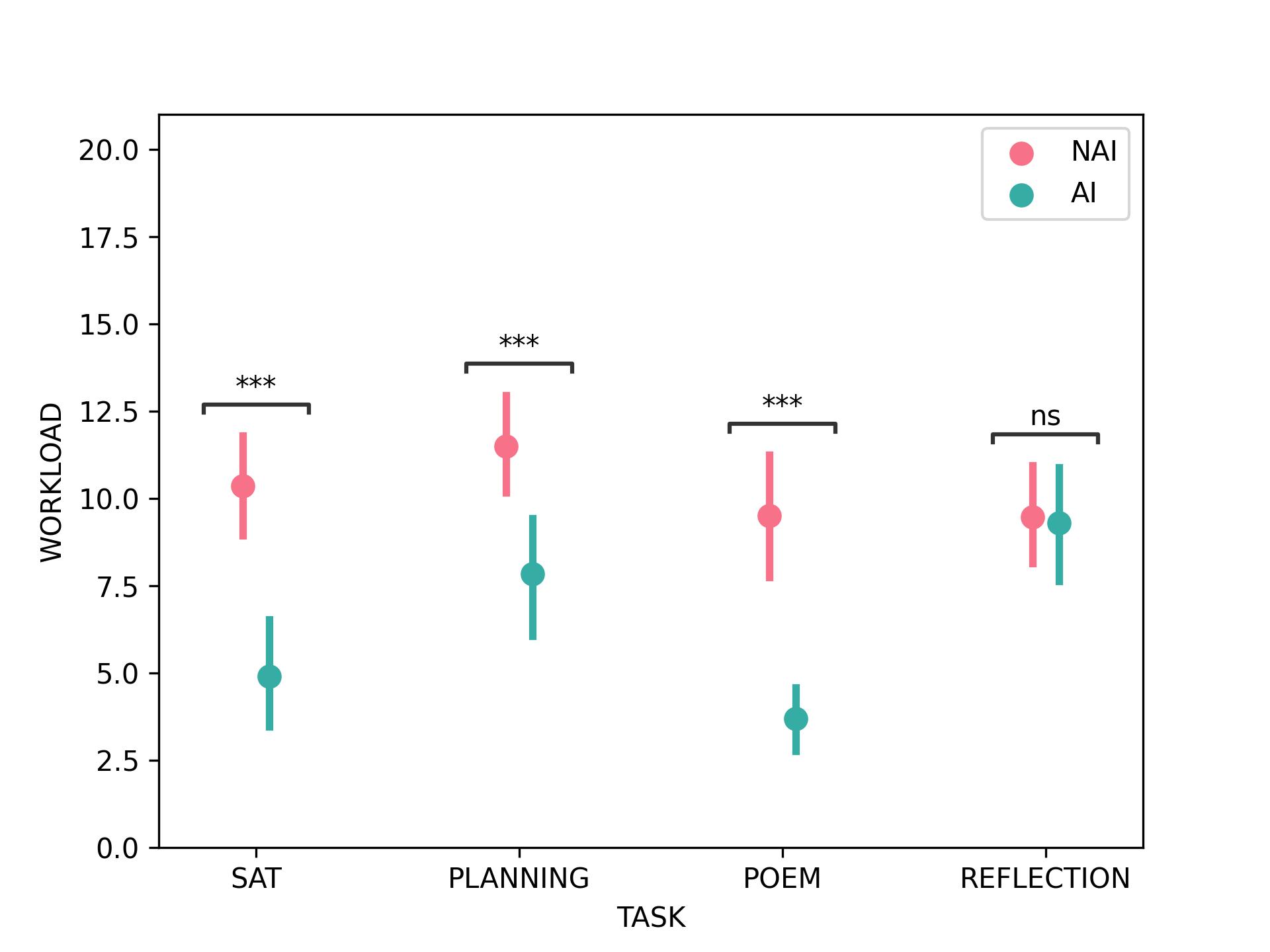}
    \caption{Self-reported workload levels were lower with Copilot for all levels of \texttt{TASK} except \texttt{REFLECTION}, which shows no change.}
    \label{fig:CONDITION_TASK}
\end{figure}

\texttt{CONDITION $\times$ TASK} demonstrated a strong effect ($F_{3, 133}=7.07, p<0.001, \epsilon_{p}^{2}=.12$). Pairwise contrasts shown in Table \ref{tab:postHoc} and visualized in Figure \ref{fig:CONDITION_TASK} show that the \texttt{AI} was significantly less than \texttt{NAI} for all levels of \texttt{TASK} with the notable exception of \texttt{REFLECTION} ($t_{133}=0.17, p=0.864, \epsilon_{p}^{2}=0.00$), which did not show a significant change. This result is as-expected in terms of decreases in workload decreases for the more objective \texttt{SAT} and \texttt{PLANNING}, and in terms of no change for \texttt{REFLECTION}, but it is somewhat surprising that the participants reported a large decrease in workload with Copilot in the more subjective \texttt{POEM}.  

\begin{table}[h!]
\caption{Effects of Copilot use on self-reported mental workload within levels of \texttt{TASK}. Copilot reduced self-reported workload for all tasks except \texttt{REFLECTION}.}
\label{tab:postHoc}
\centering
\begin{adjustbox}{width=\linewidth}
\begin{tabular}{lcrrrrrccc}
\toprule
\textbf{Task} & \textbf{Contrast} & \textbf{Est.} & \textbf{SE} & \textbf{df} & \textbf{t} & \textbf{p} & \textbf{sig.} & \textbf{$\epsilon_p^2$} & \textbf{$\epsilon_p^2$ CI} \\
\midrule
SAT        & NAI - AI & 5.46 & 0.97 & 133 & 5.63 & $<$0.001 & *** & 0.19 & [0.08, 0.30] \\
POEM       & NAI - AI & 5.80 & 0.97 & 133 & 5.98 & $<$0.001 & *** & 0.21 & [0.10, 0.32] \\
PLANNING   & NAI - AI & 3.66 & 0.97 & 133 & 3.77 & $<$0.001 & *** & 0.09 & [0.02, 0.19] \\
REFLECTION & NAI - AI & 0.17 & 0.97 & 133 & 0.17 & 0.864    & ns  & 0.00 & [0.00, 0.00] \\
\bottomrule
\end{tabular}
\end{adjustbox}
\end{table}

\subsubsection{\nameref{RQ1}-C Results}
\begin{figure}[h!]
\centering
\includegraphics[width=.8\linewidth]{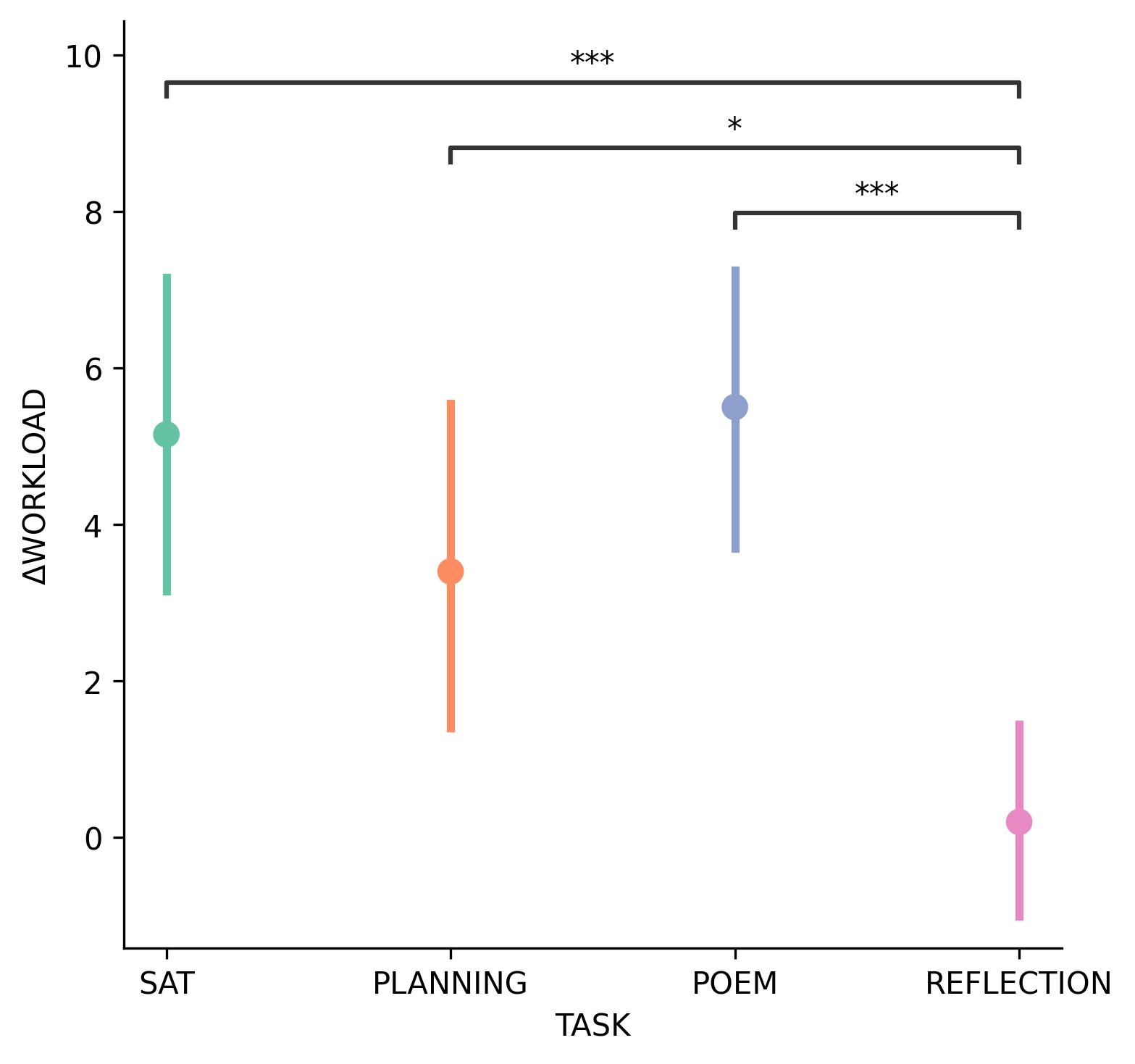}
\caption{Effect of Copilot use on self-reported workload across tasks. Larger values indicate that Copilot decreased workload by a larger amount. Self-reported TLX scores were significantly lowered by Copilot in all tasks as compared to \texttt{REFLECTION}.}
\label{fig:DELTA_TLX}
\end{figure}

Results regarding \nameref{RQ1}-C in consideration of changes due to Copilot use are shown in Figure \ref{fig:DELTA_TLX} and Table \ref{tab:postHocDelta_TLX}. Copilot significantly reduced workload in all tasks in relation to \texttt{REFLECTION}: \texttt{POEM - REFLECTION} ($t_{133}=4.11, p<0.001, \epsilon_{p}^{2}=0.11$), \texttt{SAT - REFLECTION} ($t_{133}=3.86, p<0.001, \epsilon_{p}^{2}=0.09$), and \texttt{PLANNING - REFLECTION} ($t_{133}=2.54, p=0.012, \epsilon_{p}^{2}=0.04$).

\begin{table}[h!]
\centering
\caption{Contrast results comparing the effect of \texttt{AI} versus \texttt{NAI} across levels of \texttt{TASK} on self-reported workload. The decrease in workload accounted for by Copilot was significantly larger in \texttt{SAT}, \texttt{POEM}, and \texttt{PLANNING} than in \texttt{REFLECTION}.}
\label{tab:postHocDelta_TLX}
\begin{adjustbox}{width=\linewidth}
\begin{tabular}{lcrrrrrrrr}
\toprule
\textbf{Contrast}&\textbf{Effect}&\textbf{Est.}&\textbf{SE}&\textbf{df}&\textbf{t}&\textbf{p}&\textbf{sig.}&\textbf{$\epsilon_{p}^{2}$}&\textbf{$\epsilon_{p}^{2}$ CI}\\
\midrule
POEM - REFLECTION & AI - NAI & 5.63 & 1.37 & 133 & 4.11 & $<$0.001 & *** & 0.11 & [0.03,0.21] \\
SAT - REFLECTION & AI - NAI & 5.29 & 1.37 & 133 & 3.86 & $<$0.001 & *** & 0.09 & [0.02,0.20] \\
PLANNING - REFLECTION &AI - NAI &  3.49 & 1.37 & 133 & 2.54 & 0.012 & * & 0.04 & [0.00,0.12] \\
PLANNING - POEM & AI - NAI & -2.14 & 1.37 & 133 & -1.56 & 0.121 & ns & 0.01 & [0.00,0.07] \\
PLANNING - SAT & AI - NAI & -1.80 & 1.37 & 133 & -1.31 & 0.192 & ns & 0.01 & [0.00,0.06] \\
POEM - SAT & AI - NAI & 0.34 & 1.37 & 133 & 0.25 & 0.804 & ns & 0.00 & [0.00,0.00] \\
\bottomrule
\end{tabular}
\end{adjustbox}
\end{table}

\subsubsection{\nameref{RQ1} Results Summary}
As expected, self-reported workload decreased with Copilot in relation to the gradient of subjectivity: \texttt{SAT} and \texttt{PLANNING} exhibited large decreases, whereas \texttt{REFLECTION} did not. Surprisingly, we also noted the largest overall decrease in self-reported workload during \texttt{POEM}. These results indicate that LLM-use may be helpful to users during subjective tasks which are purely creative, but not in subjective tasks which engage episodic memory. 

\subsection{\nameref{RQ2}: fNIRS Results}

\subsubsection{\nameref{RQ2}-A and \nameref{RQ2}-B Results}
Detailed results are in Table \ref{tab:ANOVA_VLF_LF}. The use of Copilot did not effect \texttt{fNIRS} for either \texttt{DSI} or \texttt{DS$\upphi$} either the left (\texttt{DSI}: $F_{1,52}=2.60, p=0.113, \epsilon_{p}^{2}=0.03$; \texttt{DS$\upphi$}: $F_{1,51.04}=2.14, p=0.150, \epsilon_{p}^{2}=0.02$) or right (\texttt{DSI}: $F_{1,52}=0.61, p=0.437, \epsilon_{p}^{2}=0.00$; \texttt{DS$\upphi$}: $F_{1,39.0}=0.17, p=0.683, \epsilon_{p}^{2}=0.00$) sides. Similarly, no effects were found among the interaction of \texttt{CONDITION}$\times$\texttt{TASK} for either measure in the left (\texttt{DSI}: $F_{1, 52}=1.25, p=0.302, \epsilon_{p}^{2}=0.01$; \texttt{DS$\upphi$}: $F_{1, 50.98}=1.90, p=0.142, \epsilon_{p}^{2}=0.05$) or right (\texttt{DSI}: $F_{1, 52}=0.54, p=0.655, \epsilon_{p}^{2}=0.00$; \texttt{DS$\upphi$}: $F_{1, 52}=0.39, p=0.736, \epsilon_{p}^{2}=0.00$). These results indicate that, despite self-reported workload changes, there were not large measurable changes in PFC activity due to differential VLFO patterns as a consequence of Copilot use. One possible explanation is that the differences in any difficulty levels between the tasks' baselines and the Copilot use was not extreme, for example as in similar levels of the N-Back task \cite{herff2014mental}.

\begin{table}[h!]
\centering
\caption{Results of modeling formula \ref{formula:initial} with \texttt{fNIRS} as the \texttt{DV} for all four combinations of [\texttt{L}, \texttt{R}], and [\texttt{DSI}, \texttt{DS$\upphi$}]. These results indicate significant activation changes in \texttt{DSI} of the right PFC based on  \texttt{TASK}. No effect on prefrontal activity in either the left or right PFC, or in relation to \texttt{DS$\upphi$}, is shown under \texttt{CONDITION}. Note that \textbf{sig.} considers adjusted $\alpha$ of 0.025, correcting across tests for \texttt{DSI} and \texttt{DS$\upphi$} with each of \texttt{L} and \texttt{R}, separately.}
\label{tab:ANOVA_VLF_LF}
\begin{adjustbox}{width=\linewidth}
\begin{tabular}{lllrrrrrrr}
\toprule
\textbf{Side}&\textbf{Meas}&\textbf{Factor}&\textbf{df1}&\textbf{df2}&\textbf{F}&\textbf{p}&\textbf{sig.}&\textbf{$\epsilon_{p}^{2}$}&\textbf{$\epsilon_{p}^{2}$ CI}\\
\midrule
L & DSI & CONDITION & 1 & 52.0 & 2.60 & 0.113 & ns & 0.03 & [0.00,0.14] \\
L & DSI & TASK & 3 & 39.0 & 1.40 & 0.256 & ns & 0.03 & [0.00,0.09] \\
L & DSI & CONDITION x TASK & 3 & 52.0 & 1.25 & 0.302 & ns & 0.01 & [0.00,0.04] \\
\midrule
L & DS$\upphi$ & CONDITION & 1 & 51.04 & 2.14 & 0.150 & ns & 0.02 & [0.00,0.13] \\
L & DS$\upphi$ & TASK & 3 & 39.08 & 1.19 & 0.330 & ns & 0.01 & [0.00,0.03] \\
L & DS$\upphi$ & CONDITION x TASK & 3 & 50.98 & 1.90 & 0.142 & ns & 0.05 & [0.00,0.13] \\
\midrule
\midrule
R & DSI & CONDITION & 1 & 52.0 & 0.61 & 0.437 & ns & 0.00 & [0.00,0.00] \\
R & DSI & TASK & 3 & 39.0 & 3.52 & 0.024 & * & 0.15 & [0.00,0.30] \\
R & DSI & CONDITION x TASK & 3 & 52.0 & 0.54 & 0.655 & ns & 0.00 & [0.00,0.00] \\
\midrule
R & DS$\upphi$ & CONDITION & 1 & 52.0 & 0.17 & 0.683 & ns & 0.00 & [0.00,0.00] \\
R & DS$\upphi$ & TASK & 3 & 39.0 & 0.20 & 0.898 & ns & 0.00 & [0.00,0.00] \\
R & DS$\upphi$ & CONDITION x TASK & 3 & 52.0 & 0.39 & 0.763 & ns & 0.00 & [0.00,0.00] \\
\bottomrule
\end{tabular}
\end{adjustbox}
\end{table}

\subsubsection{\nameref{RQ2}-C Results}
\begin{figure}[h!]
    \centering
    \includegraphics[width=\linewidth]{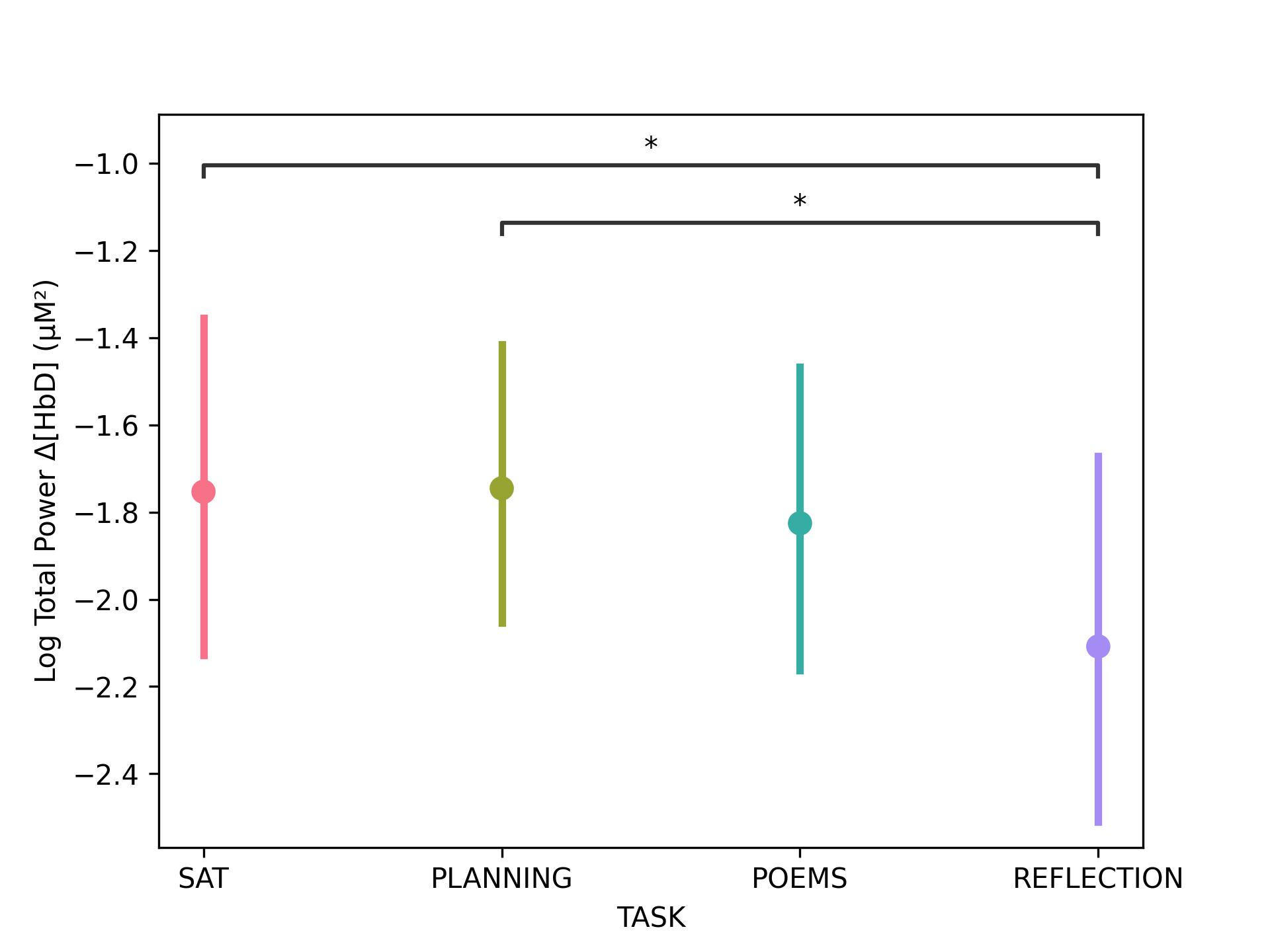}
    \caption{Log total power of $\Delta$[HbD] of the VLF band in the right prefrontal probe compared across tasks, irrespective of \texttt{CONDITION}. Note that lower total power indicates higher prefrontal activation. The \texttt{REFLECTION} task demonstrated higher levels of activation as compared to \texttt{SAT} and \texttt{PLANNING}, likely due to its engagement of episodic memory.}
    \label{fig:FNIRS_VLFO}
\end{figure}

Irrespective of \texttt{CONDITION}, \texttt{TASK} showed significance with a strong effect size as measured on the right aspect of the PFC in the \texttt{DSI} measurement ($F_{3, 39}=3.52, p=0.024, \epsilon_{p}^{2}=0.15$): post-hoc contrasts were therefore run for \texttt{TASK} within the right probe. Results are shown in Table \ref{tab:postHocCOND_VLFO_TASK}, and visualized in Figure \ref{fig:FNIRS_VLFO}. Of note are differences between \texttt{PLANNING} - \texttt{REFLECTION}  ($t_{39}=2.82, p=0.036, \epsilon_{p}^{2}=0.15$) and \texttt{SAT} - \texttt{REFLECTION} ($t_{39}=2.76, p=0.042, \epsilon_{p}^{2}=0.14$); and although not significant, given the effect size we also note \texttt{POEM} - \texttt{REFLECTION} ($t_{39}=2.20, p=0.142, \epsilon_{p}^{2}=0.09$). These results indicate a difference in PFC activity as a consequence of \texttt{TASK}, specifically indicating that the episodic memory task \texttt{REFLECTION} induced higher prefrontal cortex activation as compared to the other tasks.

\begin{table}[h!]
\centering
\caption{\texttt{VLF} $\Delta$[HbD] contrast results for the \texttt{TASK} factor. \texttt{REFLECTION} showed decreased activity in the \texttt{VLF} band, indicating increased prefrontal activation, as compared to the \texttt{SAT} and \texttt{PLANNING} tasks, irrespective of \texttt{CONDITION}. }
\label{tab:postHocCOND_VLFO_TASK}
\begin{adjustbox}{width=\linewidth}
\begin{tabular}{clrrrrrccc}
\toprule
\textbf{Side} & \textbf{Contrast} & \textbf{Est.} & \textbf{SE} & \textbf{df} & \textbf{t} & \textbf{p} & \textbf{sig.} & \textbf{$\epsilon_{p}^{2}$} & \textbf{$\epsilon_{p}^{2}$ CI} \\
\midrule
R & PLANNING - REFLECTION & 0.36 & 0.13 & 39.00 & 2.82 & 0.036 & * & 0.15 & [0.01,0.35] \\
R & SAT - REFLECTION & 0.35 & 0.13 & 39.00 & 2.76 & 0.042 & * & 0.14 & [0.01,0.35] \\
R & POEM - REFLECTION  & 0.28 & 0.13 & 39.00 & 2.20 & 0.142 & ns & 0.09 & [0.00,0.28] \\
R & PLANNING - POEM & 0.08 & 0.13 & 39.00 & 0.62 & 0.924 & ns & 0.00 & [0.00,0.00] \\
R & POEM - SAT & -0.07 & 0.13 & 39.00 & -0.56 & 0.942 & ns & 0.00 & [0.00,0.00]  \\
R & PLANNING - SAT & 0.01 & 0.13 & 39.00 & 0.06 & 1.000 & ns & 0.00 & [0.00,0.00] \\
\bottomrule
\end{tabular}
\end{adjustbox}
\end{table}

\subsubsection{\nameref{RQ2} Results Summary}
No changes were found in prefrontal activation as related to Copilot use; however, significant differences were seen across \texttt{TASK} in the right PFC: namely, between \texttt{REFLECTION} and \texttt{SAT}/\texttt{PLANNING}, which likely results from the \texttt{REFLECTION} task's engagement of episodic memory. 

\subsection{\nameref{RQ3}: Empatica Results}
To answer this we first developed separate initial models where we use each of the signals of interest as defined in section \ref{EDASymp} as the \texttt{DV} in Formula \ref{formula:initial}. Results shown in Table \ref{tab:EMPATICA}.

\begin{table}[h!]
\centering
\caption{Results from separate models created from Formula \ref{formula:initial} with each measurement type as \texttt{DV}. No physiological measurements from the Empatica E4 device showed significant changes as a consequence of \texttt{TASK}, \texttt{CONDITION}, or their interaction. Note that for \texttt{HR} and \texttt{HRV} tests $\alpha$ is set to 0.025 due to similarity of the research question underlying the tests.}
\label{tab:EMPATICA}
\begin{adjustbox}{width=\linewidth}
\begin{tabular}{llrrrrccc}
\toprule
\textbf{Measure} & \textbf{Factor} & \textbf{df1} & \textbf{df2} & \textbf{F} & \textbf{p} & \textbf{sig.} & \textbf{$\epsilon_{p}^{2}$} & \textbf{$\epsilon_{p}^{2}$ CI} \\
\midrule
\midrule
HR & CONDITION & 1 & 77 & 3.29 & 0.074 & ns & 0.03 & [0.00,0.11] \\
HR & TASK & 3 & 77 & 0.33 & 0.801 & ns & 0.00 & [0.00,0.00] \\
HR & CONDITION$\times$TASK & 3 & 77 & 0.46 & 0.712 & ns & 0.00 & [0.00,0.00] \\
\midrule
HRV & CONDITION & 1 & 56.31 & 0.34 & 0.561 & ns & 0.00 & [0.00,0.00] \\
HRV & TASK & 3 & 57.02 & 1.18 & 0.324 & ns & 0.00 & [0.00,0.00] \\
HRV & CONDITION$\times$TASK & 3 & 56.29 & 0.41 & 0.743 & ns & 0.00 & [0.00,0.00] \\
\midrule
\midrule
EDA & CONDITION & 1 & 77 & 0.03 & 0.858 & ns & 0.00 & [0.00,0.00] \\
EDA & TASK & 3 & 77 & 0.27 & 0.844 & ns & 0.00 & [0.00,0.00] \\
EDA & CONDITION$\times$TASK & 3 & 77 & 0.46 & 0.710 & ns & 0.00 & [0.00,0.00] \\
\bottomrule
\end{tabular}
\end{adjustbox}
\end{table}

\subsubsection{\nameref{RQ3}-A, \nameref{RQ3}-B, and \nameref{RQ3}-C Results}
A marginal effect with low effect size of \texttt{CONDITION} on \texttt{HR} was observed ($F_{1,77}=3.29, p=0.074, \epsilon_{p}^{2}=0.03$); no significant changes in any of the Empatica E4 measures were observed either within or across tasks. These findings suggest that stress as measured by cardiovascular and electrodermal activity is unchanged by Copilot use, tasks along the gradient of subjectivity, and the interaction of these factors. 

\subsection{\nameref{RQ4}: Quality Results}

\begin{table}[ht]
\centering
\caption{\texttt{Quality} ANOVA results. Note that, due to the varying distribution of data, \texttt{SAT} was put in a separate model from the other levels of \texttt{TASK}. \texttt{CONDITION} showed significance for \texttt{SAT}, and \texttt{CONDITION}, \texttt{TASK}, and their interaction all showed significant effects for the other model. }
\label{tab:ANOVA_QUALITY}
\begin{adjustbox}{width=\linewidth}
\begin{tabular}{llrrrrccc}
\toprule
\textbf{Model} & \textbf{Factor} & \textbf{df1} & \textbf{df2} & \textbf{F} & \textbf{p} & \textbf{sig.} & \textbf{$\epsilon_{p}^{2}$} & \textbf{$\epsilon_{p}^{2}$ CI} \\
\midrule
SAT & CONDITION & 1 & 20& 15.00 & $<$0.001 & *** & 0.40 & [0.13,0.60] \\
\midrule
OTHERS & CONDITION & 1 & 100 & 6.90 & 0.010 & ** & 0.06 & [0.01,0.14] \\
OTHERS & TASK & 2 & 100 & 7.18 & $<$0.001 & ** & 0.11 & [0.02,0.20] \\
OTHERS & CONDITION$\times$TASK & 2 & 100 & 3.53 & 0.033 & * & 0.05 & [0.00,0.12] \\
\bottomrule
\end{tabular}
\end{adjustbox}
\end{table}

\subsubsection{\nameref{RQ4}-A and \nameref{RQ4}-B Results}
\begin{figure}
    \centering
    \includegraphics[width=\linewidth]{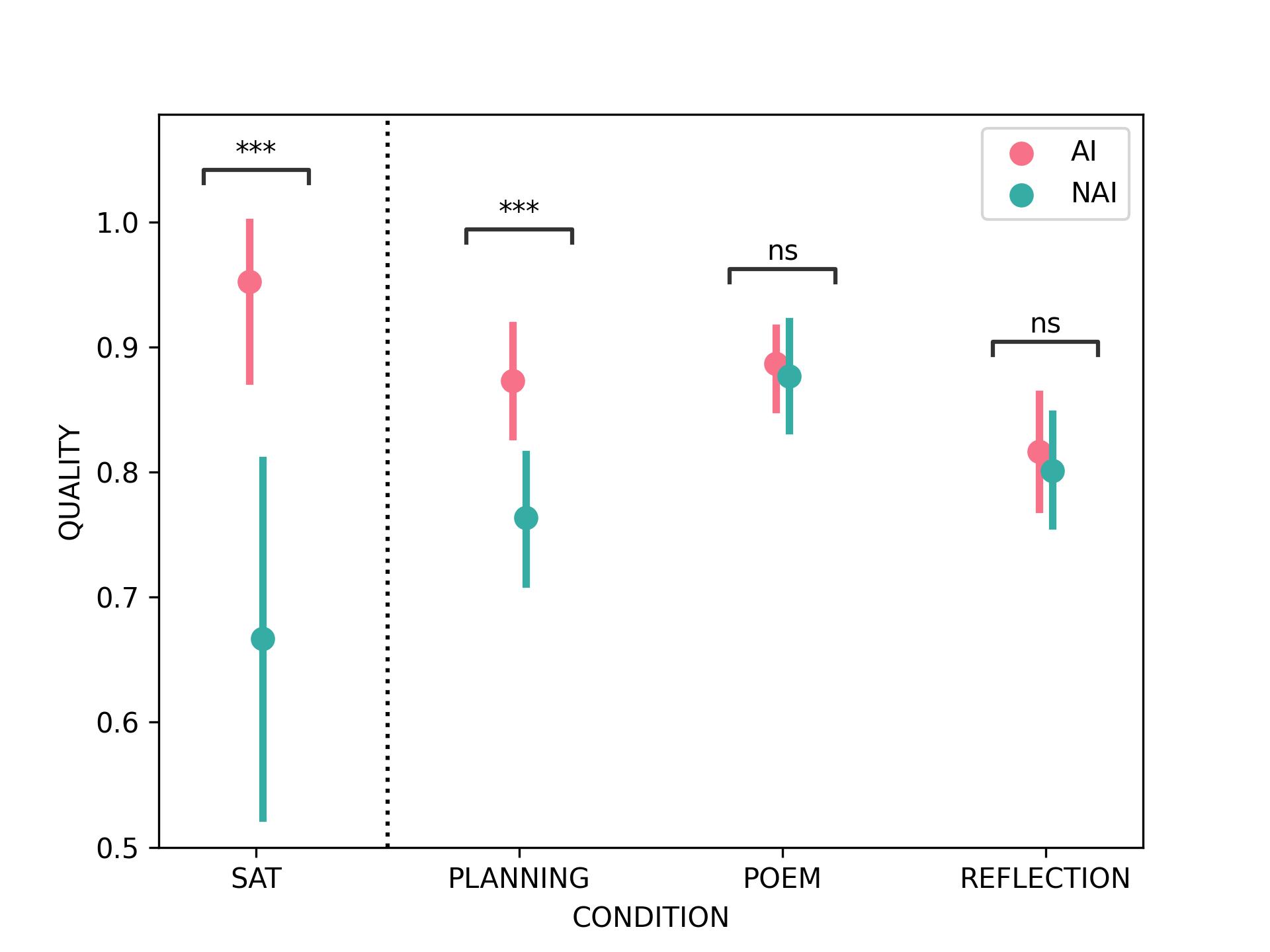}
    \caption{\texttt{SAT} and \texttt{PLANNING} tasks had significantly higher \texttt{QUALITY} scores in the \texttt{AI} condition. \texttt{POEM} and \texttt{REFLECTION} showed no change. Note that the \texttt{SAT} data was trained on a separate model because of distinctions in grading methodology. } 
    \label{fig:QUALITY_TASK}
\end{figure}

There is a significant effect of \texttt{CONDITION} for both \texttt{SAT} ($F_{1, 20}=15, p<0.001, \epsilon_{p}^{2}=.40$) and the other tasks ($F_{1,100}=6.9, p=0.01,\epsilon_{p}^{2}=0.06$). For the three other tasks there is likewise an effect of \texttt{CONDITION $\times$ TASK} ($F_{2, 100}=3.53, p<0.033, \epsilon_{p}^{2}=.05$), but Figure \ref{fig:QUALITY_TASK} and Table \ref{tab:postHocQuality} show that within these three tasks the only significant task is \texttt{PLANNING} ($t_{100}=3.68, p<0.001, \epsilon_{p}^{2}=0.11$). These results indicate an increase in \texttt{QUALITY} for the more objective tasks, but not the more subjective ones. 

\begin{table}[h!]
\centering
\caption{\texttt{QUALITY} contrast results for all levels of \texttt{TASK} excluding \texttt{SAT}. Only \texttt{PLANNING} increased in the \texttt{AI} condition as compared to the \texttt{NAI} condition. }
\label{tab:postHocQuality}
\begin{adjustbox}{width=\linewidth}
\begin{tabular}{lcrrrrrccc}
\toprule
\textbf{Task} & \textbf{Contrast} & \textbf{Est.} & \textbf{SE} & \textbf{df} & \textbf{t} & \textbf{p} & \textbf{sig.} & \textbf{$\epsilon_{p}^{2}$} & \textbf{$\epsilon_{p}^{2}$ CI} \\
\midrule
PLANNING & AI - NAI & 0.11 & 0.03 & 100 & 3.68 & $<$0.001 & *** & 0.11 & [0.02,0.23] \\
REFLECTION & AI - NAI & 0.02 & 0.03 & 100 & 0.53 & 0.600 & ns & 0.00 & [0.00,0.00] \\
POEM & AI - NAI & 0.01 & 0.03 & 100 & 0.34 & 0.735 & ns & 0.00 & [0.00,0.00] \\
\bottomrule
\end{tabular}
\end{adjustbox}
\end{table}

\subsubsection{\nameref{RQ4}-C Results} 
\begin{figure}[h!]
\centering
\includegraphics[width=.8\linewidth]{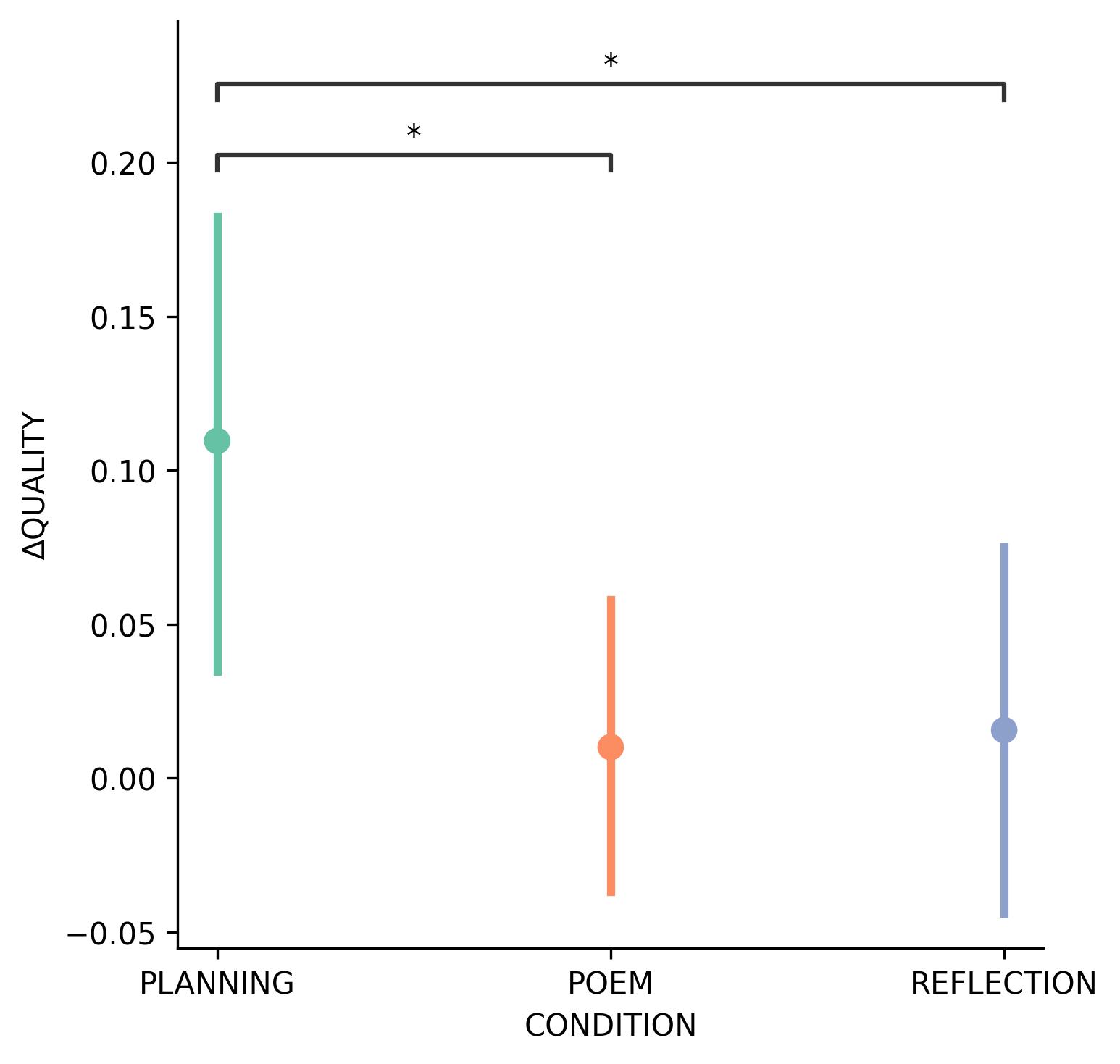}
\caption{Effect of Copilot use on \texttt{QUALITY} scores across tasks. \texttt{QUALITY} increased significantly with Copilot in the \texttt{PLANNING} as compared to \texttt{POEM} and \texttt{REFLECTION}.}
\label{fig:QUALITY_DELTA}
\end{figure}

The largest effect is seen in \texttt{SAT}. Post-hoc contrasts observing effects across tasks for the changes between quality of \texttt{AI} versus \texttt{NAI}, shown in Table \ref{tab:postHocDeltaQuality} and visualized in Figure \ref{fig:QUALITY_DELTA}, showed that the effect of the increase in \texttt{QUALITY} score of Copilot use is significantly higher in \texttt{PLANNING} as compared to \texttt{POEM} ($t_{100}=2.36, p=0.020, \epsilon_{p}^{2}=0.04$) and \texttt{REFLECTION} ($t_{100}=2.23, p=0.028, \epsilon_{p}^{2}=0.04$). These results indicate that Copilot may be beneficial in terms of quality output for more objective tasks. 

\begin{table}[ht]
\centering
\caption{Contrast results comparing the effect of \texttt{AI} versus \texttt{NAI} across levels of TASK on Quality scores. The increase in quality accounted for by Copilot was larger in \texttt{PLANNING} than in \texttt{POEM} or \texttt{REFLECTION}.}
\label{tab:postHocDeltaQuality}
\begin{adjustbox}{width=\linewidth}
\begin{tabular}{lcrrrrrccc}
\toprule
\textbf{Contrast}&\textbf{Effect}&\textbf{Est.}&\textbf{SE}&\textbf{df}&\textbf{t}&\textbf{p}&\textbf{sig.}&\textbf{$\epsilon_{p}^{2}$}&\textbf{$\epsilon_{p}^{2}$ CI}\\
\midrule
PLANNING - POEM & AI - NAI & 0.10 & 0.04 & 100 & 2.36 & 0.020 & * & 0.04 & [0.00,0.14] \\
PLAN - REFLECTION & AI - NAI & 0.09 & 0.04 & 100 & 2.23 & 0.028 & * & 0.04 & [0.00,0.14] \\
POEM - REFLECTION & AI - NAI & -0.01 & 0.04 & 100 & -0.13 & 0.896 & ns & 0.00 & [0.00,0.00] \\
\bottomrule
\end{tabular}
\end{adjustbox}
\end{table}

\subsubsection{ICC Results} Scores for \texttt{OVERALL} ($\text{ICC}=0.774, \text{95\% CI}=[0.7, 0.83]$), \texttt{PLANNING} ($\text{ICC}=0.817, \text{95\% CI}=[0.69, 0.9]$), \texttt{REFLECTION} ($\text{ICC}=0.751, \text{95\% CI}=[0.58, 0.86]$), and \texttt{POEM} ($\text{ICC}=0.652, \text{95\% CI}=[0.42, 0.8]$) were all moderate. Within this range, however, we observed the expected behavior regarding our \texttt{ICC} measurement in that the more open-ended and subjective tasks demonstrated lower consistency scores, with the 95\% lower CI for the \texttt{POEM} task rating as poor. 

\subsubsection{\nameref{RQ4} Results Summary}
In summary, \texttt{QUALITY} scores for \texttt{SAT} and \texttt{PLANNING} increased with Copilot use, and the increase in quality score with Copilot use significantly differed between \texttt{PLANNING} and \texttt{POEM}/\texttt{REFLECTION}. These results indicate that for more objective tasks, Copilot use can increase \texttt{QUALITY}, whereas for more subjective tasks, it is less likely to do so.  

\subsection{\nameref{RQ5}: Enjoyment Results - Quantitative Evaluation}
\label{RQ5 Quantitative Evaluation Results}

\subsubsection{\nameref{RQ5}-A and \nameref{RQ5}-B Results} 
\begin{figure}[h!]
    \centering
    \includegraphics[width=\linewidth]{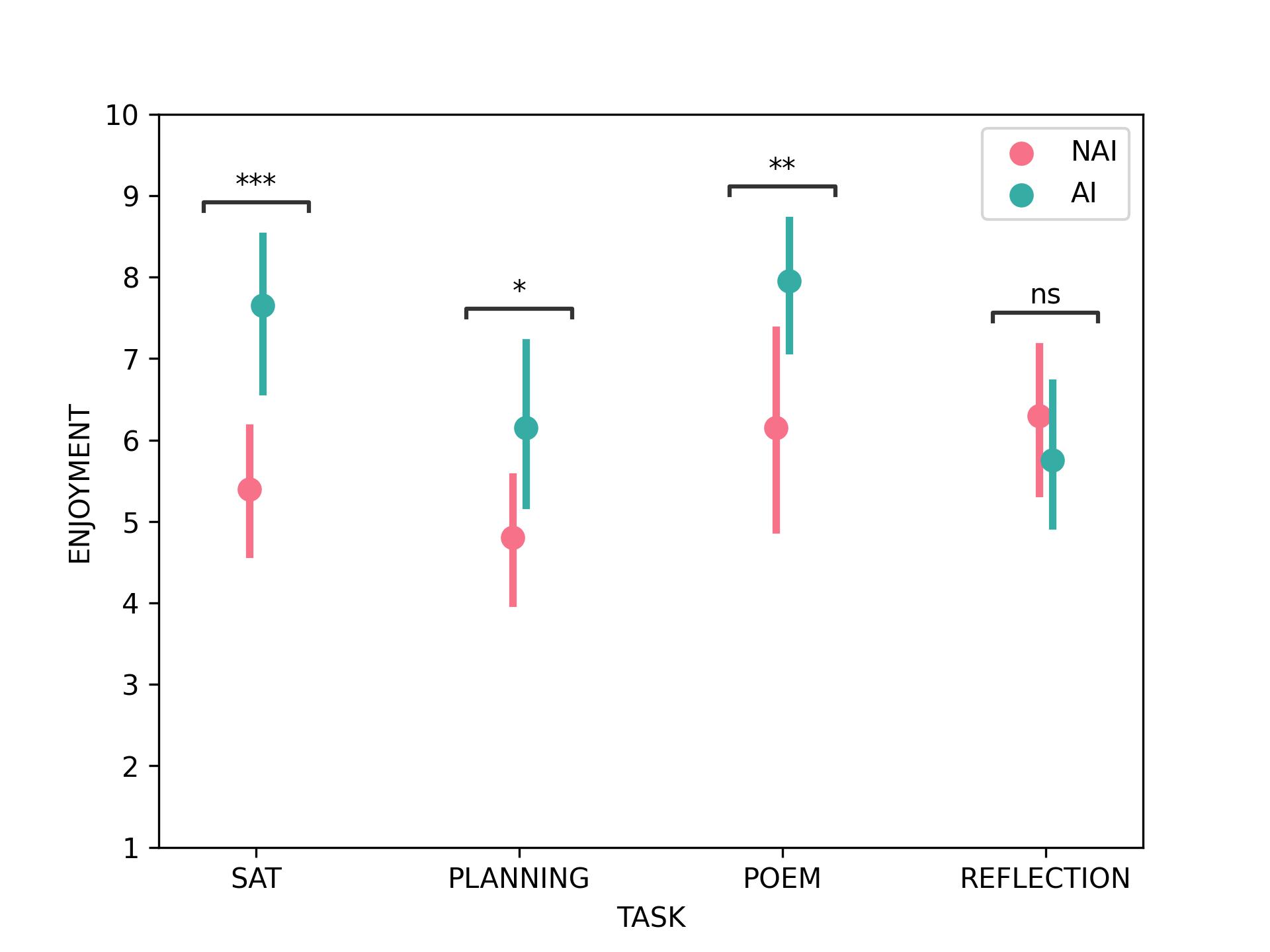}
    \caption{\texttt{ENJOYMENT} between \texttt{CONDITION} across \texttt{TASK}. While \texttt{SAT} and \texttt{POEM} demonstrated increases in \texttt{ENJOYMENT} with Copilot, no change was found for \texttt{PLANNING} or \texttt{REFLECTION}.}
    \label{fig:enjoyment_condition_task}
\end{figure}

See Table \ref{tab:ENJOYMENT}. Participants reported higher \texttt{ENJOYMENT} when using Copilot ($F_{1,133}=15.06, p<0.001, \epsilon_{p}^{2}=0.05$). Contrast results (see Table \ref{tab:postHocENJOY} and Figure \ref{fig:enjoyment_condition_task}) indicate that, with the exception of \texttt{REFLECTION} ($t_{133}=-0.88, p=0.380, \epsilon_{p}^{2}=0.00$), this is likewise true for each individual task. These results directly parallel the self-reported results for \texttt{TLX}, and indicate that, in addition to objective tasks, participants enjoyed using the Copilot assistant for subjective tasks which did not require self-reflection, and did not enjoy its use during reflective tasks.

\begin{table}[h!]
\centering
\caption{ANOVA results of Formula \ref{formula:initial} with \texttt{ENJOYMENT} as the \texttt{DV}; significant results were found for \texttt{CONDITION}, \texttt{TASK}, and their interaction.}
\label{tab:ENJOYMENT}
\begin{adjustbox}{width=\linewidth}
\begin{tabular}{lrrrrccc}
\toprule
\textbf{Factor}&\textbf{df1} &\textbf{df2}&\textbf{F}&\textbf{p}&\textbf{p.sig}&\textbf{$\epsilon_{p}^{2}$}&\textbf{ $\epsilon_{p}^{2}$ CI}\\
\midrule
CONDITION & 1 & 133 & 15.06 & $<$0.001 & *** & 0.10 & [0.03,0.18] \\
TASK & 3 & 133 & 4.66 & $<$0.001 & ** & 0.07 & [0.01,0.14] \\
CONDITION$\times$TASK & 3 & 133 & 3.88 & 0.01 & * & 0.06 & [0.00,0.12] \\
\bottomrule
\end{tabular}
\end{adjustbox}
\end{table}

\begin{table}[h!]
\centering
\caption{Contrast results of \texttt{ENJOYMENT} (\texttt{AI}-\texttt{NAI}) within levels of \texttt{TASK}. All levels showed significant increases in \texttt{ENJOYMENT} during \texttt{AI}, with the notable exception of \texttt{REFLECTION}, which showed no significant change.}
\label{tab:postHocENJOY}
\begin{adjustbox}{width=\linewidth}
\begin{tabular}{lcrrrrrccc}
\toprule
\textbf{Task} & \textbf{Contrast} & \textbf{Est.} & \textbf{SE} & \textbf{df} & \textbf{t} & \textbf{p} & \textbf{sig.} & \textbf{$\epsilon_{p}^{2}$} & \textbf{$\epsilon_{p}^{2}$ CI} \\
\midrule
SAT & AI - NAI & 2.25 & 0.62 & 133 & 3.60 & $<$0.001 & *** & 0.08 & [0.02,0.18] \\
POEM & AI - NAI & 1.80 & 0.62 & 133 & 2.88 & 0.005 & ** & 0.05 & [0.00,0.14] \\
PLANNING & AI - NAI & 1.35 & 0.62 & 133 & 2.16 & 0.033 & * & 0.03 & [0.00,0.10] \\
REFLECTION & AI - NAI & -0.55 & 0.62 & 133 & -0.88 & 0.380 & ns & 0.00 & [0.00,0.00] \\
\bottomrule
\end{tabular}
\end{adjustbox}
\end{table}

\subsubsection{\nameref{RQ5}-C Results}
\begin{figure}[h!]
\centering
\includegraphics[width=.8\linewidth]{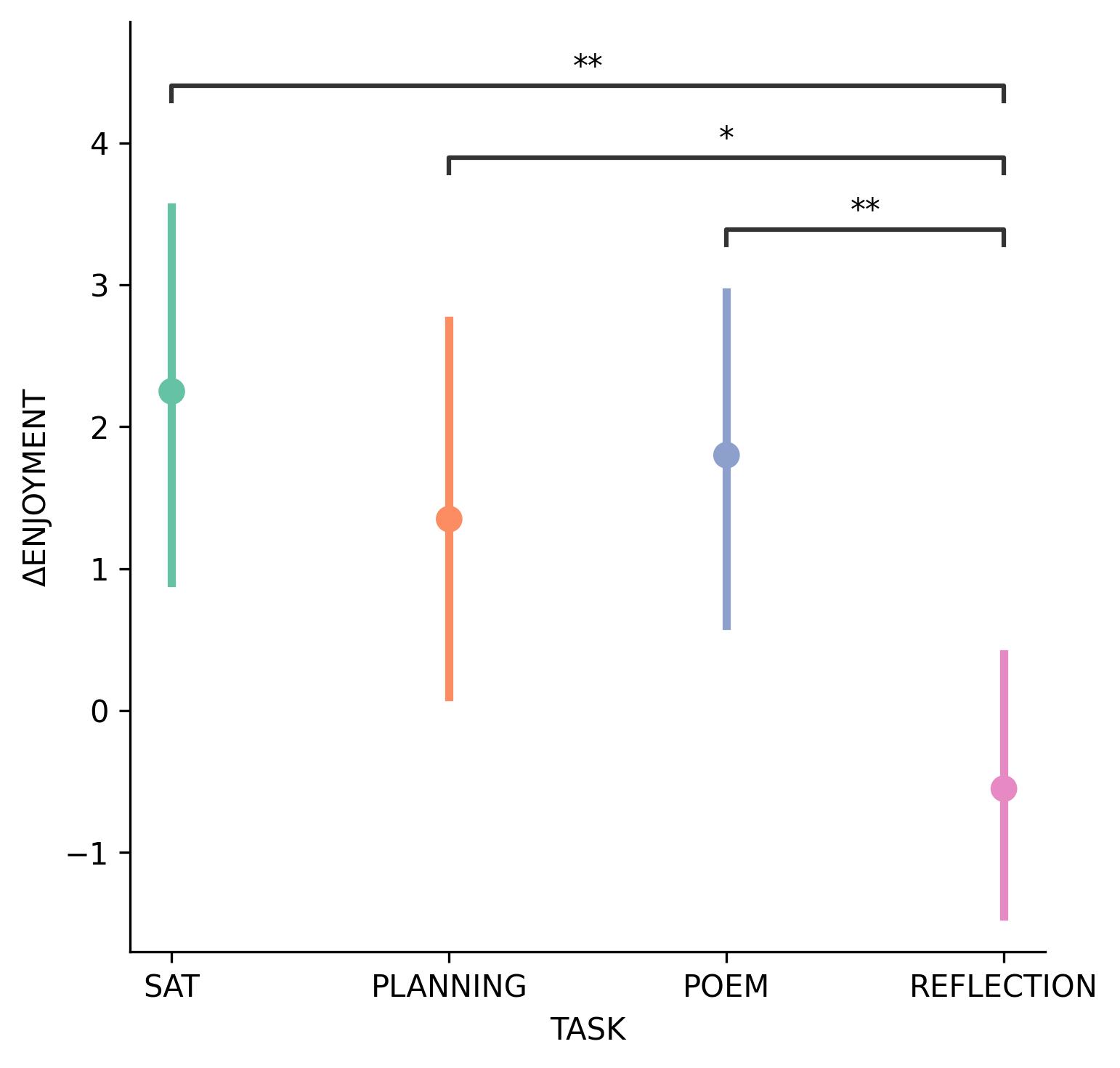}
\caption{Effect of Copilot on \texttt{ENJOYMENT} scores compared across \texttt{TASK}. Similar to the changes in \texttt{TLX}, \texttt{ENJOYMENT} increased significantly with Copilot in the all tasks as compared \texttt{REFLECTION}.}
\label{fig:POST_HOC_DELTA_ENJOY}
\end{figure}

Results are shown in Table \ref{tab:postHocDELTA_ENJOY} and Figure \ref{fig:POST_HOC_DELTA_ENJOY}: change in self-reported enjoyment with Copilot was higher for the all of the tasks as compared to \texttt{REFLECTION} (\texttt{SAT} - \texttt{REFLECTION}: $t_{133}=3.17, p=0.002, \epsilon_{p}^{2}=0.06$; \texttt{POEM} - \texttt{REFLECTION}: $t_{133}=2.66, p=0.009, \epsilon_{p}^{2}=0.04$, \texttt{PLANNING} - \texttt{REFLECTION}: $t_{133}=2.15, p=0.033, \epsilon_{p}^{2}=0.03$). 

\subsubsection{RQ5 Results Summary} 
These results mirror those of \texttt{TLX}, indicating that, although Copilot provided tangible benefits both in the purely objective tasks (\texttt{SAT}, \texttt{PLANNING}) as well in a creative task (\texttt{POEM}), it did not have any benefits during the episodic memory task (\texttt{REFLECTION}). 

\begin{table}[h!]
\caption{Contrast results comparing \texttt{AI} versus \texttt{NAI} across \texttt{TASK}. All levels of \texttt{TASK} showed higher \texttt{ENJOYMENT} in \texttt{AI} versus \texttt{NAI} as compared to \texttt{REFLECTION}.}
\label{tab:postHocDELTA_ENJOY}
\begin{adjustbox}{width=\linewidth}
\begin{tabular}{lcrrrrrrccc}
\toprule
\textbf{Contrast}&\textbf{Effect}&\textbf{Est.}&\textbf{SE}&\textbf{df}&\textbf{t}&\textbf{p}&\textbf{sig.}&\textbf{$\epsilon_{p}^{2}$}&\textbf{$\epsilon_{p}^{2}$ CI}\\
\midrule
SAT - REFLECTION & AI - NAI &  2.80 & 0.88 & 133 & 3.17 & 0.002 & ** & 0.06 & [0.01,0.16] \\
POEM - REFLECTION & AI - NAI & 2.35 & 0.88 & 133 & 2.66 & 0.009 & ** & 0.04 & [0.00,0.13] \\
PLANNING - REFLECTION & AI - NAI & 1.90 & 0.88 & 133 & 2.15 & 0.033 & * & 0.03 & [0.00,0.10] \\
PLANNING - SAT & AI - NAI & -0.90 & 0.88 & 133 & -1.02 & 0.310 & ns & 0.00 & [0.00,0.03] \\
PLANNING - POEM & AI - NAI & -0.45 & 0.88 & 133 & -0.51 & 0.611 & ns & 0.00 & [0.00,0.00] \\
POEM - SAT & AI - NAI & -0.45 & 0.88 & 133 & -0.51 & 0.611 & ns & 0.00 & [0.00,0.00] \\
\bottomrule
\end{tabular}
\end{adjustbox}
\end{table}

\subsection{\nameref{RQ5}: Enjoyment Results - Qualitative Evaluation}
\label{RQ5 Qualitative Evaluation Results}
After each task, users were asked to write optional comments response to their overall experience with the task and the usefulness of the AI tool.

\subsubsection{Reading comprehension} As expected, most users found Copilot exceptionally helpful in completing the SAT reading comprehension questions. LLMs perform well with highly structured tasks such as reading a passage and answering multiple choice questions about it. However, not all users trusted that Copilot would be accurate, with user 3 stating that ``\textit{I would not want to use the AI tool for such a task because I feel like I would then not put in the effort of checking if the answers given are correct and then I would later on be in self doubt about whether or not the answers were correct}''. This lack of trust reduced the likelihood that they might benefit from access to an LLM, even for tasks in which the tool shines. 

\subsubsection{Planning} User 19 succinctly puts it: ``\textit{the AI helps a lot with idea generation that can be worked on}'', essentially saying that Copilot was especially helpful in generating ideas and content that could then be refined by the user. However, as other users found, in order to benefit from the generative capabilities of the LLM, a basic understanding of its functionality was necessary. User 12 found that ``\textit{the tool refuses to look up specific information I requested and repeatedly came back with generic responses despite being asked to  `be specific'. It was more frustrating than helpful after adopting its initial response as I end up combating with AI to get the information I want}''.

\subsubsection{Poem} Most people had little experience writing poems or didn't like writing them, meaning that Copilot was especially useful in helping them complete the task given the strict time constraints. However, some users felt that they were of a lower quality, with user 15 stating that ``\textit{Having AI for this task was helpful but made the whole ordeal quite boring and the poem, in the end, was not representative of my own feelings and emotions. While it was easier, I did feel like using AI for this kind of assignment yields quite ordinary pieces of work}''.

\subsubsection{Reflection} Similar to the poem task, users found that Copilot was ineffective in helping them write about their personal experiences and feelings in relation to art. However, one unique advantage the LLM tool provided was the ability to access information when writing the personal reflection, with user 10 finding that ``\textit{The tool definitely helped in giving a brief introduction to the album which would have required additional research on my part}''.

\subsubsection{Trends} These comments reveal that Copilot was especially helpful in a generative capacity, creating drafts or providing information that could then be refined when completing the task. However, multiple factors mitigated the potential benefits of Copilot: a lack of trust in Copilot's answers, a lack of understanding of its functionality, difficulties with iterating on content, and its inability to interact with or produce personal content. Many users also felt that the time lag between prompting the tool and receiving a response diminished the system's usefulness. 

\section{Discussion} 
\subsubsection{Self-reported \texttt{WORKLOAD}, \texttt{QUALITY}, and \texttt{ENJOYMENT}} Regarding self-reported measures, Copilot's overall effect on users was as-expected for the objective tasks within the gradient of subjectivity: with Copilot, users reported decreased \texttt{TLX} workload and increased \texttt{ENJOYMENT} in \texttt{SAT} and \texttt{PLANNING}; this was coupled with increases in \texttt{QUALITY}. On the opposing end of the subjectivity gradient we likewise found expected results: for \texttt{REFLECTION}, participants reported no tangible changes as a consequence of Copilot use, nor was there a measured change in \texttt{PERFORMANCE}. 

Compared to the other results, \texttt{POEM} produced a set of somewhat unexpected findings: namely, a large decrease in \texttt{TLX} workload coupled with an increase in \texttt{ENJOYMENT}. Were initially surprised with these results given the high degree of subjectivity in \texttt{POEM}. However, based on user comments, we believe that this result is partially due to the fact that our users were not used to writing poems; that is, Copilot's ability to produce a significant quantity of reasonable output nearly instantly made the task both easier and more enjoyable. This finding mirrors other work that has indicated that AI-related tools provide the most benefit to the least experienced users \cite{Butler2023Microsoft}. Given that the participants were novice poetry writers, we would caution extrapolation of this finding to the full set of creative domains, and encourage follow-up studies exploring the population of creative users in more depth. Further, no change in output quality was observed in \texttt{POEM}.

\subsubsection{fNIRS}
Given the decreases in \texttt{TLX} workload for three of the tasks when using Copilot, we were slightly surprised to see a disparity in terms of no findings in the \texttt{fNIRS} data to a similar regard. Of note, however, is that although our study tasks certainly required users' effort, none of them required an \textit{extreme} amount of mental workload (along the lines of the NBack task, for instance \cite{herff2014mental}); that is, tasks which require higher levels of mental effort under the baseline condition may be necessary in order to distinguish levels of prefrontal cortex activation as reflected in VLFO measurements.

A notable finding was an increase in activation of the right PFC during \texttt{REFLECTION} as compared to \texttt{SAT} and \texttt{PLANNING}, irrespective of Copilot use. This result likely stems from the \texttt{REFLECTION} task's engagement of different underlying psycho-physiological state: that of self-reflection and autobiographical episodic memory retrieval. As discussed earlier, these states have been shown to increase prefrontal activation \cite{Gilbert2006FunctionalSpecialization}, and specifically have been linked to right prefrontal activation \cite{nolde1998role, tulving1994hemispheric}. And more broadly, self-reflection, self-referential states, and episodic memory activation have been linked to the larger Default Mode Network (DMN) \cite{menon202320}. Thus, in conjunction with the \texttt{TLX}, \texttt{QUALITY}, and \texttt{ENJOYMENT} results, the neural finding implies that the helpfulness of AI assistants decreases in response to increased levels of activation of episodic memory; it is also possible that this link is related more broadly to DMN activation. 

\subsubsection{Other Physiological Results}
Given that there were no significant effects related to \texttt{HR}, \texttt{HRV}, or \texttt{EDA}, we can conclude that neither the effects of Copilot use, nor tasks across the gradient of subjectivity, are extreme in the physiological domain outside of the brain.

\section{Conclusion}
We tested Copilot, an interactive LLM-based AI assistant, using a multimodal set of measurement techniques including prefrontal cortex activation via fNIRS in terms of its effects on user states through a variety of tasks designed along a gradient of subjectivity intended to become increasingly difficult for the assistant. Results indicate that for tasks which are challenging yet tightly constrained overall in terms of objectivity, users benefit in terms of decreases in self-reported mental workload and increases in reported enjoyment and objective performance. For creative tasks for new users with more subjective criteria for success (\texttt{POEM}), Copilot produced very similar gains to the more objective tasks, despite our expectations; however, these results should be interpreted with caution as participants may not have approached this purely creative task with the same level of rigor as the others. In purely reading-comprehension tasks (\texttt{SAT}), the distinction between neural activation as measured by fNIRS was not statistically significant. Lastly, we found that Copilot was not able to assist users meaningfully in tasks which require primarily subjective material (\texttt{REFLECTION}), and that brain measurement via fNIRS indicated larger prefrontal cortex activation during this task than the others, likely due to episodic memory retrieval and potentially DMN activation. We concretely specify the activation of neural states related to episodic memory as a shortcoming of artificial agents, and more tentatively indicate that the lack of the assistant's ability to help users may align with a broader activity of the DMN. While this is an initial study with a single LLM-based AI tool, more will be required in the domain of evaluation of effects of AI assistants on human users. 

\section{Acknowledgments}
Kenny, Soraya, Microsoft, Brent Hecht, Darren Gergle

\bibliographystyle{IEEEtran}
\bibliography{references}

\include{supp_mat}

\end{document}

%% file: supp_mat.tex
\Huge{Supplementary Material}
\section{Planning Tasks}
\normalsize
\label{APPENDIX:TASK_LIST}

\subsection{Planning Task A:  Future Leaders Retreat}
Construct a short (½ - 1 page) plan for a "Future Leaders Retreat" intended for emerging student leaders from REDACTED University. This retreat will focus on personal leadership development, resilience training, and introspection. Ensure that your plan includes: 
\begin{enumerate}
    \item A reflective name for the retreat that resonates with personal growth. 

    \item Agenda highlights such as mindfulness sessions, personal leadership journey sharing, and resilience building workshops.  

    \item A specific serene location (on or off campus) conducive to introspection and inner growth. 

    \item Considerations required for the holistic development and well-being of the attendees.  

    \item Plan for candidate selection for the retreat.  
\end{enumerate}

\subsection{Planning Task B: Alumni Leadership Summit: REDACTED University Elite Networking Event}
Draft a short (½ - 1 page) plan for an exclusive business networking event targeting REDACTED University alumni in leadership positions. Your plan should specify: 
\begin{enumerate}
    \item A dynamic event name that signifies industry leadership and networking.  

    \item Keynote speakers of interest, industry panel discussions, and insights into business trends.  

    \item A location near or on REDACTED University that embodies a business-centric environment.  

    \item Strategies to promote inter-industry networking and engagement between alumni and ambitious students.  

    \item Note that you may pick an area of expertise for the summit which relates to your field of study (or possible majors for you if undecided). 
\end{enumerate}

\section{Poetry Tasks}
\subsection{Poetry Task A: Nature}
Write a brief (10–15 line) poem on the beauty of nature. 
\subsection{Poetry Task B: Joy}
Imagine a moment of unexpected joy on an ordinary day. Write a short (10-15 line) poem capturing the essence of that emotion.  

\section{Reflection Tasks}
\subsection{Reflection Task A: Movie}
Pick your favorite movie released before 2020. Then draft a 2-paragraph reflection on how the movie resonates with your personal experiences or memories. Use as much detail as possible (quotes, scenes, etc). 

\subsection{Reflection Task B: Album} 
Pick your favorite album released before 2020. Then draft a 2-paragraph reflection on how the album resonates with your personal experiences or memories. Use as much detail as possible (song lyrics, album themes, etc).

\section{SAT Tasks}
The SAT tasks were slightly modified version of the 2016 SAT practice tests: numbers 5 \cite{CollegeBoard2016_alcazar} and 7 \cite{CollegeBoard2016_sat_gold}.

\section{Gradient of Subjectivity: Potential Confound Analysis}
\label{APPENDIX:CONFOUND_SUBTASK}

\section{Subtask Difficulty}

\begin{figure}
    \centering
    \includegraphics[width=.75\linewidth]{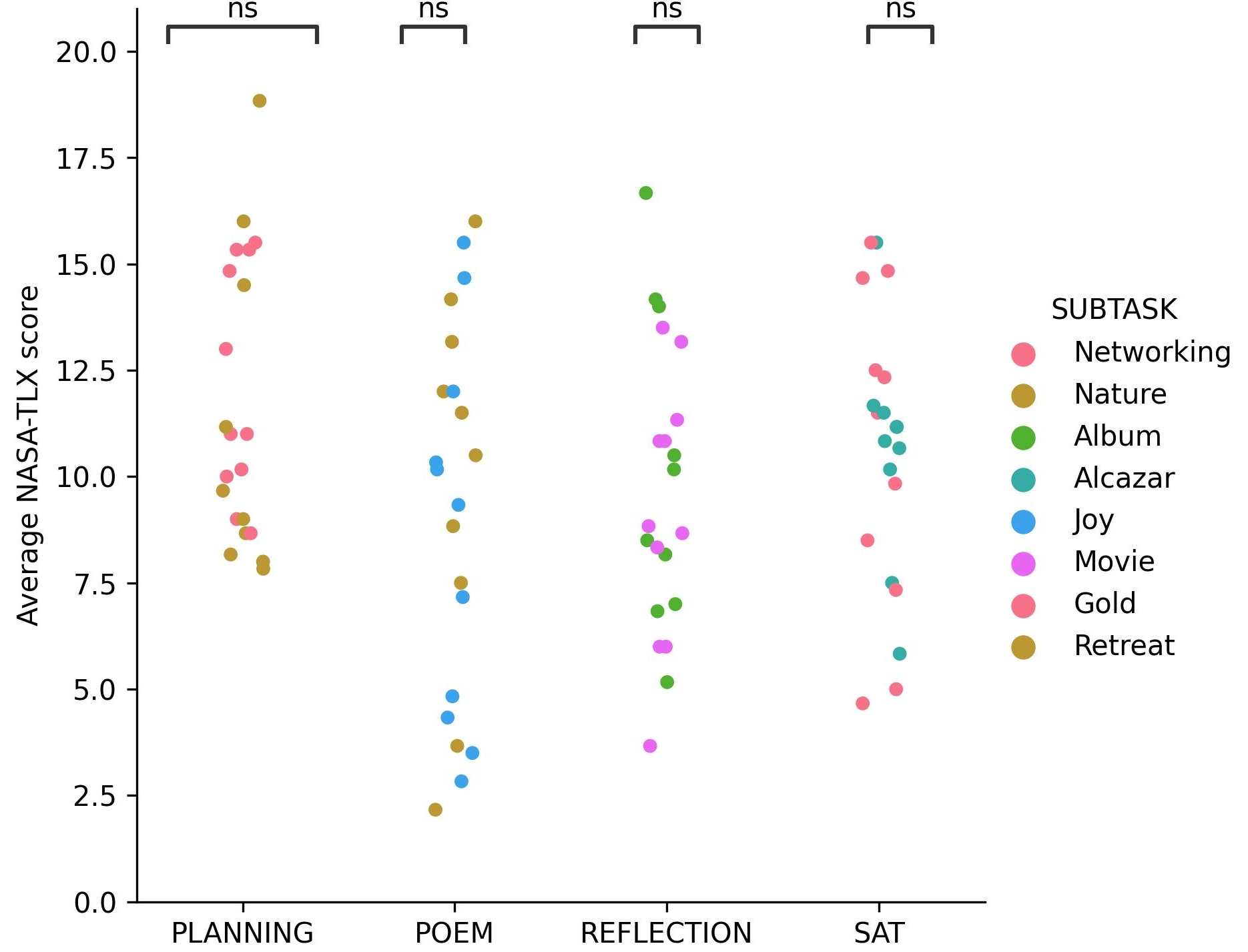}
    \caption{\texttt{NASA-TLX} Mental Workload Score within each \texttt{SUBTASK}. Within each \texttt{TASK}, none of the \texttt{SUBTASK}s were significantly more difficult than the other. }
    \label{fig:CONFOUND_SUBTASK}
\end{figure}

For a given \texttt{TASK}, although we randomized whether \texttt{SUBTASK} \texttt{A} or \texttt{B} would be done with the Copilot assistant, it is neverthless important to determine whether or not the \texttt{SUBTASK}s for each \texttt{TASK} were of equal difficulty. To do this, we analyzed the data of only the \texttt{NAI} \texttt{CONDITION} in a between-subjects manner (as each subject only did each subtask once). Specifically, we performed independent-samples t-tests for each pair of subtasks. Results are listed in Table \ref{tab:CONFOUND_SUBTASK} and Figure \ref{fig:CONFOUND_SUBTASK}. No significant results were found, indicating that the \texttt{SUBTASK}s within each \texttt{TASK} were of similar difficulty.

\begin{table}[ht]
\centering
\caption{T-Test results for \texttt{SUBTASK} Difficulty Comparison}
\label{tab:CONFOUND_SUBTASK}
\begin{tabular}{lcccc}
\hline
TASK & Df & t-value & p.adj & sig. \\\hline
POEM &  19 & -0.693 & 0.497 & ns  \\
REF &  19 & -0.615 & 0.546 &  ns \\
SAT & 19  & 0.004 & 0.997 & ns \\
PLAN &  19 & -0.690 & 0.506 & ns  \\
\hline
\end{tabular}
\end{table}

\section{Task Time}
\begin{figure}
    \centering
    \includegraphics[width=.75\linewidth]{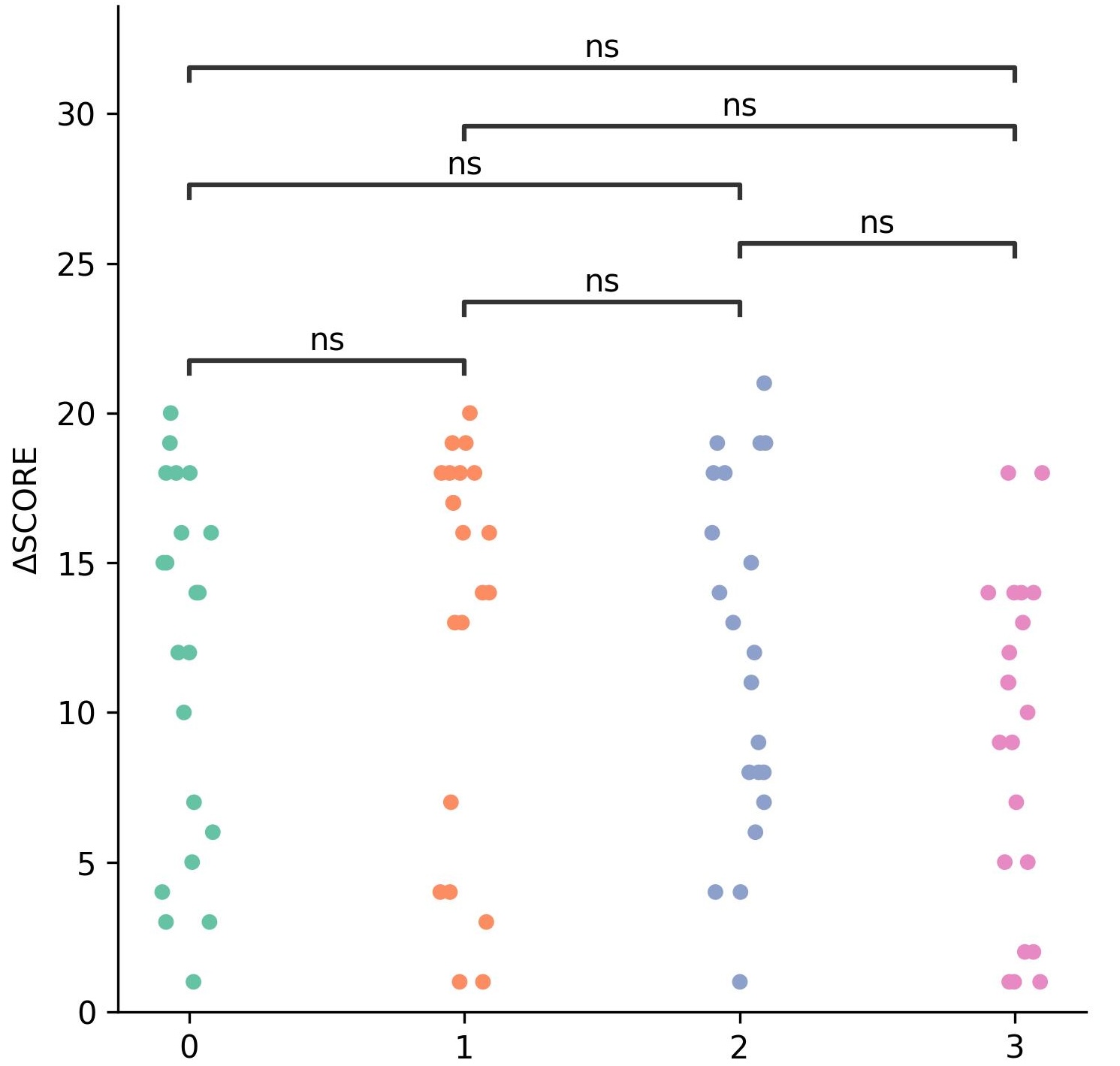}
    \caption{Change in Workload Score (\texttt{NAI} - \texttt{AI}) as a Function of Task Number}
    \label{fig:DELTA_SCORE_TASK_NUM}
\end{figure}

We also analyzed the potential confound of task time as it relates to mental workload. Specifically we were concerned that the task number would effect the change in workload scores between the \texttt{AI} and \texttt{NAI} levels of \texttt{CONDIITON}. To test this, we created a lmer model with the formula 
\begin{equation}
\Delta SCORE \sim TASK\_NUM + (1|pid)
\end{equation}
Where \texttt{TASK\_NUM} was a number from 1-4, and \texttt{$\Delta$SCORE} is the \textit{change} in score defined as \texttt{NAI} - \texttt{AI}. The ANOVA for this model did not report a significant result ($F_{3,60}$=1.87, p=0.144, $\eta_{p}^{2}$=0.09, 95\% CI=[0.00, 1.0]), although there was a moderate effect size. 
Contrast results are shown in Table \ref{tab:postHocTASKNUM} and Figure \ref{fig:DELTA_SCORE_TASK_NUM}. None of the contrasts demonstrated significance. 

\begin{table}[ht]
\centering
\caption{Post-Hoc Contrast Results for \texttt{TASK\_NUM}}
\label{tab:postHocTASKNUM}
\resizebox{\columnwidth}{!}{
\begin{tabular}{lrrrrrlrl}
Contrast & Estimate & SE & df & t.ratio & p.value & p.sig & $\eta_{p}^{2}$ & 95\% CI \\
\hline
task\_num1 - task\_num3 & 3.76 & 1.66 & 60.00 & 2.26 & 0.119 & ns & 0.08 & [0.0,1.0] \\
task\_num2 - task\_num3 & 2.81 & 1.66 & 60.00 & 1.69 & 0.339 & ns & 0.05 & [0.0,1.0] \\
task\_num0 - task\_num3 & 2.62 & 1.66 & 60.00 & 1.57 & 0.401 & ns & 0.04 & [0.0,1.0] \\
task\_num0 - task\_num1 & -1.14 & 1.66 & 60.00 & -0.69 & 0.902 & ns & 0.01 & [0.0,1.0] \\
task\_num1 - task\_num2 & 0.95 & 1.66 & 60.00 & 0.57 & 0.940 & ns & 0.01 & [0.0,1.0] \\
task\_num0 - task\_num2 & -0.19 & 1.66 & 60.00 & -0.11 & 0.999 & ns & 0.00 & [0.0,1.0] \\
\hline
\end{tabular}
}
\end{table}

